\documentclass[10pt,preprint2]{aastex}

\begin{document}

%%%%%%%%%
% Title %
%%%%%%%%%

\title{The Ensemble Photometric Variability of $\sim 25000$ Quasars in
  the Sloan Digital Sky Survey}

%%%%%%%%%%%
% Authors %
%%%%%%%%%%%

\author{
Daniel E. Vanden Berk\altaffilmark{1},
Brian C. Wilhite\altaffilmark{2},
Richard G. Kron\altaffilmark{2,3},
Scott F. Anderson\altaffilmark{4},
Robert J. Brunner\altaffilmark{5},
Patrick B. Hall\altaffilmark{6,7},
\v{Z}eljko Ivezi\'{c}\altaffilmark{6},
Gordon T. Richards\altaffilmark{6},
Donald P. Schneider\altaffilmark{8},
Donald G. York\altaffilmark{2,9},
Jonathan V. Brinkmann\altaffilmark{10},
Don Q. Lamb\altaffilmark{2},
Robert C. Nichol\altaffilmark{11},
David J. Schlegel\altaffilmark{6}
}

\altaffiltext{1}{Department of Physics and Astronomy, University of Pittsburgh, 3941 O'Hara Street, Pittsburgh, PA 15260.}
\altaffiltext{2}{Department of Astronomy and Astrophysics, The University of Chicago, 5640 South Ellis Avenue, Chicago, IL 60637.}
\altaffiltext{3}{Fermi National Accelerator Laboratory, P.O. Box 500, Batavia, IL 60510.}
\altaffiltext{4}{Department of Astronomy, University of Washington, Box 351580, Seattle, WA 98195.}
\altaffiltext{5}{Department of Astronomy/NCSA, University of Illinois, 1002 W. Green Street, Urbana, IL 61801.}
\altaffiltext{6}{Princeton University Observatory, Peyton Hall, Princeton, NJ 08544.}
\altaffiltext{7}{Departamento de Astronom\'{\i}a y Astrof\'{\i}sica, Pontificia Universidad Cat\'{o}lica de Chile, Casilla 306, Santiago 22, Chile.}
\altaffiltext{8}{Department of Astronomy and Astrophysics, The Pennsylvania State University, 525 Davey Laboratory, University Park, PA 16802.}
\altaffiltext{9}{Enrico Fermi Institute, The University of Chicago, 5640 South Ellis Avenue, Chicago, IL 60637.}
\altaffiltext{10}{Apache Point Observatory, P.O. Box 59, Sunspot, NM 88349.}
\altaffiltext{11}{Department of Physics, Carnegie Mellon University, 5000 Forbes Avenue, Pittsburgh, PA 15232.}

%%%%%%%%%%%%
% Abstract %
%%%%%%%%%%%%

\begin{abstract}
Using a sample of over 25000 spectroscopically confirmed quasars
from the Sloan Digital Sky Survey, we show how quasar variability in
the rest frame optical/UV regime depends upon rest frame time lag,
luminosity, rest wavelength, redshift, the presence of radio and
X-ray emission, and the presence of broad absorption line systems.
Imaging photometry is compared with three-band spectrophotometry
obtained at later epochs spanning time lags up to about two years.
The large sample size and wide range of parameter values allow
the dependence of variability to be isolated as a function of
many independent parameters.  The time dependence of variability
(the structure function) is well-fit by a single power law with
an index $\gamma = 0.246 \pm 0.008$, on timescales from days to
years.  There is an anti-correlation of variability amplitude with
rest wavelength -- e.g.\ quasars are about twice as variable at
$1000${\AA} as $6000${\AA} -- and quasars are systematically bluer
when brighter at all redshifts.  There is a strong anti-correlation
of variability with quasar luminosity -- variability amplitude
decreases by a factor of about four when luminosity increases by a
factor of 100.  There is also a significant positive correlation of
variability amplitude with redshift, indicating evolution of the
quasar population or the variability mechanism.  We parameterize
all of these relationships.  Quasars with RASS X-ray detections are
significantly more variable (at optical/UV wavelengths) than those
without, and radio loud quasars are marginally more variable than
their radio weak counterparts.  We find no significant difference
in the variability of quasars with and without broad absorption
line troughs.  Currently, no models of quasar variability address
more than a few of these relationships.  Models involving multiple
discrete events or gravitational microlensing are unlikely by
themselves to account for the data.  So-called accretion disk
instability models are promising, but more quantitative predictions
are needed.  \end{abstract}


%%%%%%%%%%%%%%%%
% Introduction %
%%%%%%%%%%%%%%%%

\section{Introduction}
The luminosities of quasars and other active galactic nuclei
(AGNs) have been observed to vary from X-ray to radio wavelengths,
and on time scales from several hours to many years.  The majority
of quasars exhibit continuum variability on the order of $10\%$ on
timescales of months to years.  A minority of AGNs, broadly classified
as blazars, vary much more dramatically on much shorter timescales.
The mechanisms behind quasar variability are not known, although
in principle variability is a powerful means of constraining
models for the energy source of AGNs.  The most promising models
(for non-blazar variability) include accretion disk instabilities
\citep[e.g.][]{rees84, kawaguchi98}, so-called Poissonian processes
such as multiple supernovae \citep[e.g.][]{terlevich92} or star collisions
\citep{courvoisier96, torricelli00}, and gravitational microlensing
\citep[e.g.][]{hawkins93}.  Only recently have the various models become
quantitative enough for meaningful comparison with observations.
A consensus on the observational trends with variability is emerging,
but disagreements remain and even the most fundamental relationships
need better characterization.

Several dozen studies of quasar optical broadband variability
have appeared in the literature.  A number of the more important
studies are summarized in tabular form by \citet{helfand01}
and \citet{giveon99}.  Most ensemble studies have focused on
establishing correlations between variability (defined in various
ways as a measure of the source brightness change) and a number of
parameters, most importantly time lag, quasar luminosity, rest frame
wavelength, and redshift.  Characteristic timescales of variability
range from months to years \citep[e.g.][]{collier01, cristiani96,
diclemente96, smith95, hook94, trevese94}.  The amplitude of
variability rises quickly on those timescales, but may slow or even
level off on longer timescales.

An anti-correlation between quasar variability and luminosity was reported
by \citet{angione72}, and confirmed in numerous subsequent studies
\citep{uomoto76, pica83, lloyd84, obrien88, hook94, trevese94,
cid96, cristiani96, cristiani97, paltani97, giveon99, garcia99,
hawkins00, webb00}.  Such an anti-correlation is expected in Poissonian
models, although complex versions are necessary to explain the diversity
of the relationship among quasars \citep{cid00}.

There is strong evidence from multiwavelength observations of quasars
that variability increases with decreasing rest wavelength, which
holds over a wavelength range spanning at least the ultraviolet to
near infrared \citep{cutri85, neugebauer89, kinney91, paltani94,
diclemente96, cristiani97, giveon99, cid00, helfand01, trevese02}.
The wavelength dependence is related to the observed
tendency for quasar spectra to become harder (bluer) in bright phases
\citep{cutri85, edelson90, giveon99, cid00, trevese01}. The
chromatic nature of quasar variability is often taken as evidence
against gravitational microlensing as the primary cause of variability
(e.g. \citet{cristiani97}; except see \citet{hawkins96}), although
this may be accounted for if regions closer to the center are both
brighter and bluer.

A correlation of variability with redshift is often reported
\citep{cristiani90, giallongo91, hook94, trevese94, cid96, cristiani96,
trevese02} if wavelength and luminosity dependencies are not taken into
account.  For a fixed observer timescale, the increase of variability
with increasing redshift would contradict the expected $1+z$ effect
of time dilation.  However, it has been shown by \citet{giallongo91},
\citet{cristiani96}, and \citet{cid96} that the inverse wavelength
dependence can easily account for the uncorrected redshift correlation,
since for a fixed passband in the observer frame, quasars with higher
redshifts are detected at shorter wavelengths, which systematically
vary at a greater amplitude.  It is still not clear whether any
redshift dependence remains after accounting for rest wavelength
and luminosity (which is strongly correlated with redshift in
flux-limited quasar samples).  Some studies which have leverage in
both redshift and luminosity suggest a weak correlation of redshift
and variability \citep{hook94, cristiani96}, but others show no such
effect \citep{cimatti93, paltani94, netzer96, cristiani97, helfand01}.

Variability is sometimes found to be correlated with radio loudness
\citep{pica83, smith95, garcia99, eggers00, helfand01, enya02},
the equivalent width of the H\,$\beta$ line \citep{giveon99,
cid00}, and the presence of broad absorption line troughs
\citep{sirola98}, although the results are not conclusive.  No large
X-ray detected quasar sample has been systematically studied
for optical variability, but since most blazars are X-ray bright,
a greater degree of variability may be
expected from such a sample \citep[e.g.][]{ulrich97}.

In this paper we present results on a quasar variability program
using data from the Sloan Digital Sky Survey \citep[SDSS,][]{york00}.
A complementary variability study by \citet{devries03} presents
a comparison of the SDSS Early Data Release \citep{stoughton02}
imaging photometry with archival photographic plate data.  One of
the goals of the present work is to characterize the spectroscopic
calibrations of the SDSS, in order to examine the spectroscopic
variability properties of quasars and other objects observed in
the spectroscopic survey.  The present work uses the broad band
fluxes of the spectra convolved with the SDSS filter transmission
functions in direct comparison with the imaging photometry.
This provides photometric data at two epochs in three bands for
every spectroscopically confirmed quasar in the survey -- a sample
size currently of over 25000 quasars.  This is by far the largest
quasar UV/optical variability study to date, and it also includes
the largest samples of radio selected, X-ray selected, and broad
absorption line quasars ever examined for variability.  Our goal is
to characterize the ensemble dependence of variability on many quasar
parameters and types, on timescales from weeks up to several years.

We describe the quasar sample drawn from the SDSS in
\S\,\ref{dataset}.  Ensemble measurements of the variability are
given in \S\,\ref{measure}.  We disentangle the dependence of
variability upon time lag, luminosity, wavelength, and redshift in
\S\,\ref{primary}, show how quasar colors change with variability
in \S\,\ref{color}, and look at variability in various quasar
subclasses in \S\,\ref{secondary}.  The implications of the
results are discussed in \S\,\ref{discussion}, and we conclude in
\S\,\ref{conclusions}.
% In a companion paper \citep{ivezic03} we
% examine the photometric variability properties of quasars and other
% objects from overlapping imaging runs taken at different epochs.
Throughout the paper we assume a flat, cosmological constant
dominated cosmology with parameter values $\Omega_\Lambda = 0.7,
\Omega_{M} = 0.3,$ and $H_{0}=65$km/s/Mpc.

%%%%%%%%%%%%%%%
% The Dataset %
%%%%%%%%%%%%%%%

\section{The Quasar Dataset\label{dataset}}

%
% The SDSS
%
\subsection{The Sloan Digital Sky Survey\label{SDSS}}

The Sloan Digital Sky Survey (SDSS) is a project to image $10^{4}$
deg$^{2}$ of sky mainly in the northern Galactic cap, in five broad
photometric bands ($u,g,r,i,z$) to a depth of $r \sim 23$, and to
obtain spectra of $10^{6}$ galaxies and $10^{5}$ quasars observed in
the imaging survey \citep{york00}.  All observations are made with a
dedicated 2.5m telescope at Apache Point Observatory in New Mexico.
Images are taken with a large mosaic CCD camera \citep{gunn98}
in a drift-scanning mode.  Absolute astrometry for point sources
is accurate to better than $100$ milliarcseconds \citep{pier03}.
Site photometricity and extinction monitoring are carried out
simultaneously with a dedicated 20in telescope at the observing site
\citep{hogg01}.   The imaging data are reduced and calibrated using
the PHOTO software pipeline \citep{lupton01}.  In this study we use
the point-spread function (PSF) magnitudes, which are determined
by convolving the reduced imaging data with a model of the spatial
point-spread function.  The PSF magnitudes are more stable than
aperture magnitudes for point sources, since they are less dependent
upon seeing variations, and because the PSF background noise is
less within the survey seeing limit (which is 1.7 arcseconds).
The SDSS photometric system is normalized so that the $u,g,r,i,z$
magnitudes are on the AB system \citep{smith02,fukugita96,oke83}.
The photometric zeropoint calibration is accurate to better than
$1\%$ (root-mean-squared) in the $g$, $r$, and $i$, bands, and to
better than $2\%$ in the $u$ and $z$ bands, measured by comparing
the photometry of objects in scan overlap regions.  The SDSS image
reduction and calibration routines have evolved throughout the course
of the survey and the imaging runs have been reprocessed accordingly.
Thus the object imaging magnitudes deemed ``best'', and which we
use in this study, may be slightly different than those used for
the spectroscopic target selection, although any differences are
insignificant to the results of this study.  Throughout this paper
we use magnitudes corrected for Galactic extinction according to
\citet{schlegel98}.

Objects are selected for spectroscopic follow-up as candidate
galaxies \citep{strauss02,eisenstein01}, quasars \citep{richards02},
and stars \citep{stoughton02}.  The spectroscopic targets are
grouped by 3 degree diameter areas or ``tiles'' \citep{blanton03}.
For each tile, an aluminum plate is drilled with holes corresponding
to the sky locations of the targets along with holes for blank
sky, calibration stars, and guide stars.  The plates are placed
at the focal plane of the telescope, and optical fibers run from
the hole positions to two spectrographs, each of which accepts 320
fibers allowing for the simultaneous observation of 640 objects.
For each plate, approximately 500 galaxies, 50 quasars, and 50
stars are observed.  Spectroscopic observations generally occur up
to a few months, but occasionally years, after the corresponding
imaging observations, depending upon scheduling constraints.
The spectroscopic data for this study come from 479 spectroscopic
plates observed and processed through September 2002; 284 of the plates
are part of the SDSS First Data Release \citep[DR1,][]{abazajian03},
publicly available since April 2003.  Seven of the (291) DR1 plates are
not included in this study, since the DR1 plate list was not finalized
until after the sample for this study had been gathered.

%
% The Quasar Sample
%
\subsection{Quasar Target Selection and Sample Definition\label{sample}}

Quasar candidates are selected from the imaging sample by
their non-stellar colors from the five-band photometry as
well as by matching SDSS point sources with FIRST radio sources
\citep{stoughton02}.  The selection is similar to that described by
\citet{richards02}, but the formal implementation of this algorithm
was imposed after the cutoff date for the DR1 quasar sample, and
much of the post-DR1 data used in this study.  About two-thirds of
the candidates are confirmed to be quasars from the spectroscopic
survey.  Ultraviolet excess quasars are targeted to a limit of $i
= 19.1$ and higher redshift quasars are targeted to $i = 20.2$.
These criteria give a sample that is estimated to be over 90\%
complete \citep{richards02}.  Additional quasars are targeted as
part of the SERENDIPITY and ROSAT classes \citep{stoughton02} or
(incorrectly) as stars.

Quasars are identified from their spectra using a combination of
both automated classification (about $94\%$) and manual inspection
(about $6\%$) of those objects flagged by the spectroscopic pipeline
as being less reliably identified.  For the purposes of this study,
we define ``quasar'' to mean any extragalactic object with broad
emission lines (pipeline measured full width at half maximum velocity
width of $\gtrsim 1000$km${\rm s^{-1}}$) regardless of luminosity.
The definition thus includes objects which are often classified as
less luminous types of active galactic nuclei (AGNs) rather than
quasars, and excludes AGNs without strong broad emission lines
such as BL~Lacs and some extreme broad absorption line quasars.
To assemble our sample, we extract relevant data for all point
sources from the SDSS database.  Data from only one imaging and
one spectroscopic epoch are used per object to avoid giving extra
weight to any object.  Extended sources are rejected because
they complicate the spectrophotometric recalibration (\S\,\ref{calib}),
and their spectra are likely to be seriously contaminated by host
galaxy light.  Objects with bad spectra (defined to be those with
significantly long unprocessable portions of spectrum) are rejected.
Those remaining objects spectroscopically identified as stars are
used to refine the spectroscopic calibration (see \S\,\ref{calib}).
The remaining 25710 objects identified to be quasars become part
of the quasar variability sample.

A catalog of quasars found in the SDSS DR1 dataset is described by
\citet{schneider03}.  Of the 16713 listed DR1 quasars, 14705 are
included in this study.  The difference can be attributed to two
factors.  First, the definitions of ``quasar'' are slightly different
-- \citet{schneider03} impose an absolute magnitude limit of
$M_i \le -22.0$, and they make no distinction between point-like 
and extended sources.  Second, data from only 284 of the 291 DR1
spectroscopic plates are included in this study.  Over half ($57\%$)
of the quasars in this study are contained in the DR1 sample, and we
expect that our results on quasar variability would be similar (albeit
more noisy) for the full DR1 sample, except possibly for the extended
sources.

Absolute magnitudes in the rest frame $i$ band $M_{i}$, are calculated
for each quasar using dereddened observed $i$ band PSF magnitudes
and assuming a power law spectral energy distribution $f_\lambda
\propto \lambda^{\alpha_{\lambda}}$, with a wavelength slope of
$\alpha_\lambda=-1.5$.  Detailed K-corrections are not yet available
for quasars in the SDSS photometric system.  However, estimated
K-corrections using the composite spectrum from \citet{vandenberk01}
are consistent with the simple power law assumption, and the
differences are usually no greater than 0.1 magnitude at any redshift.
The differences are not significant here since the data are coarsely
binned (by statistical necessity) when we examine absolute magnitude
trends.

%
% Spectroscopy
%
\subsection{SDSS Spectroscopy and Its Calibration\label{spectro}}

Spectra are obtained in three to four consecutive 15-minute
observations.  There are 32 sky fibers, 8 spectrophotometric
standard stars and 8 reddening standards stars observed on each
plate to help with calibration of the remaining 592 science targets.
Spectral reductions and calibrations are done using the SDSS SPECTRO 2D
pipeline \citep{stoughton02}.  The 8 spectrophotometric calibration
stars are chosen to approximate the standard F0 subdwarf star $BD+17\degr4708$,
and are used by the 2D pipeline for absolute spectral flux calibration
and dereddening due to Galactic extinction.  The 2D pipeline
also calculates synthetic spectroscopic magnitudes by convolving the
calibrated spectra with SDSS $g$, $r$, and $i$ filter transmission curves,
assuming 1.2 airmasses of extinction (the spectra do not cover the
entire wavelength ranges of the $u$ and $z$ bands).  By using these
synthetic spectral magnitudes, we obtain a second photometric data
point for every spectroscopically observed quasar.  Additionally,
a signal-to-noise ($S/N$) parameter is calculated for each of the
three bands by convolving the spectral error with the transmission
curves and dividing that into the corresponding convolved flux.
The spectral magnitude signal-to-noise is essential for characterizing
magnitude difference uncertainties (see \S\,\ref{calib}).

%
% Refined Spectrophotometry
%
\subsection{Refinement of Spectroscopic Calibration \label{calib}}

\begin{figure}
\plotone{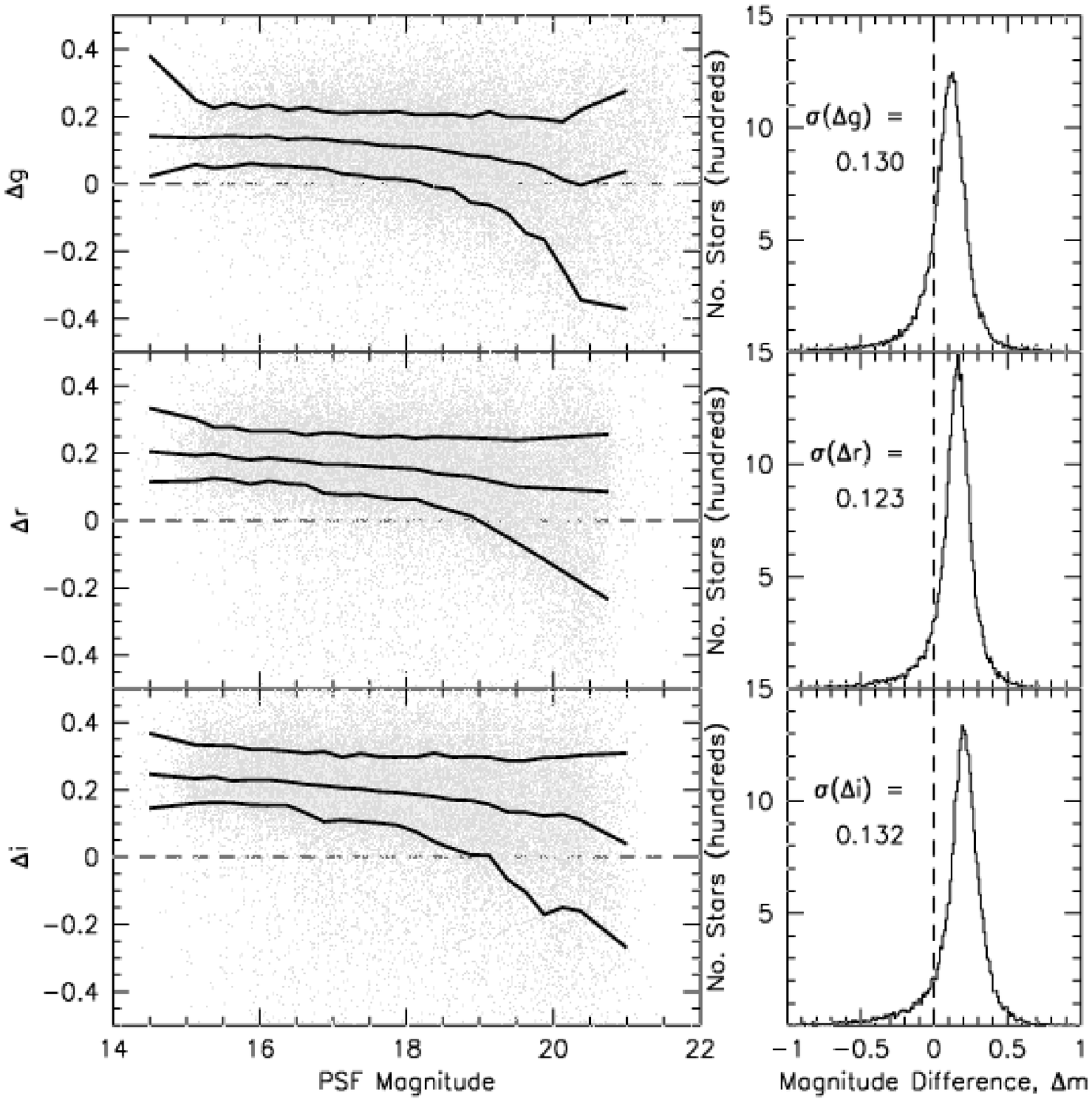}
\caption{The uncorrected spectroscopic minus photometric magnitudes 
vs. imaging PSF magnitudes for stars observed on the same
spectral plates as the quasars (left).  Results for all three bandpasses
are shown.  The curves show the binned median trends and the upper and
lower $68.3\%$ confidence envelopes.  The right side panels show the
uncorrected magnitude difference histograms.  The $68.3\%$
confidence half-widths, $\sigma$, are given for a spectroscopic S/N of 10.
\label{Fig2.1}}
\end{figure}

The differences between the spectroscopic and imaging PSF magnitudes
$\Delta m = m_{s} - m_{p}$ for all of the spectroscopically
confirmed stars in each of the $g$, $r$, and $i$ bands are shown in
Fig.\,\ref{Fig2.1}.  Well calibrated data should center around zero
magnitude difference, and there should be no trend with magnitude,
except for larger uncertainties at fainter magnitudes.  There are
clearly systematic differences in the magnitudes derived from
the spectra and from the images.  While the initial spectroscopic
calibration is more than adequate for the primary purposes of the
SDSS, namely object identification and reliable redshift measurement,
variability studies require more careful calibrations.  Fortunately,
the magnitude difference trends with PSF magnitude can be easily
understood and are almost entirely correctable.

There are three primary sources for the magnitude difference
discrepancy: the inclusion of objects with bad PSF magnitudes,
an aperture effect relating the finite fiber diameters to the PSF
magnitudes, and what may be a very small but significant sky
under-subtraction in the spectroscopic data.  Occasionally, point
sources in the images can have poor photometry if they are closely
blended with other objects, occur where the seeing has changed very
rapidly, or lie where there may be other problems in the imaging
data.  The long tails in the histograms of Fig.\,\ref{Fig2.1} are
populated mainly by the measurements of these objects.  Because the
objects will have unusual measured colors they are sometimes selected
as high-redshift quasar candidates for spectroscopic follow-up,
which turn out to be normal stars upon examination of the spectra.
Therefore, spectroscopically confirmed stars which were selected as
high-redshift quasar candidates are removed from the stellar data
set for spectrophotometric refinement.  Late-type stars identified
by the spectroscopic pipeline are also rejected because they are
often variable.

The median $\Delta{m}$ offset from zero is simply an aperture effect
wherein the $3\arcsec$ spectroscopic fibers subtend a smaller fraction
of the total object image than the point-spread function used to 
measure the PSF magnitudes in the imaging data.  The spectroscopic
fiber flux density to PSF flux density ratio is nearly a constant
(but somewhat dependent upon seeing at the spectroscopic epoch,
see below), so the magnitude difference will also be a nearly
(non-zero) constant. 

The downward trend of $\Delta m$ with PSF magnitude seen for each
band is most easily accounted for by a small overestimation of
the flux density in the  spectroscopic data, possibly caused by a slight
under-subtraction of the sky level.  Tests of the imaging photometric 
calibration show that the effect is not likely to be caused by
sky over-subtraction in the imaging data.  Further tests will have to be
done to determine the cause with certainty.  The correction for a
flux density overestimation combined with a fiber aperture
correction give
\begin{eqnarray}
  \Delta m & = & m_{s} - m_{p} = -2.5\log(f_{s}/f_{p}) \nonumber\\
           & = & -2.5\log(a(f_{p} + b)/f_{p}) \nonumber\\
           & = & -2.5\log(a + a b 10^{(m_{p} - C)/2.5})\phn,
\end{eqnarray}
where $a$ is the aperture correction, $b$ is the correction for
the flux density offset, and $C$ is the zeropoint
constant used in converting flux density to magnitude. Assuming
$a$ and $b$ are constants, the function has two adjustable parameters,
and fits to the data provide reasonably good descriptions of
the $\Delta m$ vs.\ $m_p$ trend.  However, in order to account
for any other effects, expected or unexpected, we use the
following more flexible 3 parameter function
\begin{eqnarray}
  \Delta m = {\Delta m}_{0} - \exp((m_{p} - m_{0})/m^{*})\phn,
\end{eqnarray}
where ${\Delta m}_{0}$, $m_{0}$, and $m^{*}$ are constants to
be determined from the fits to the data.  For example, the
Malmquist bias \citep{malmquist24} will add to the magnitude difference
approximately as the square of the PSF magnitude uncertainty,
${\sigma_p}^{2}$.  Since $\sigma_p$ ranges from about 0.01 to
0.05, the Malmquist bias is expected to affect the magnitude
difference by at most a few percent.  In the absence of this
or other higher order effects, the 3-parameter function would
almost exactly reproduce the 2 parameter logarithmic fit.
The 3-parameter functions are fit to the data in each band
separately, then subtracted from the magnitude differences.

After this correction, offsets from zero remain for the mean $\Delta
m$ values for stars on the same plate, in excess of those expected
from statistical uncertainty.  These plate-to-plate offsets are
due to differences in the spectral energy distributions of the
stars used as initial spectrophotometric calibrators relative to
$BD+17\degr4708$ and due to differences in the seeing at the epochs of
the plate observations.  The former effect applies to ``half plates''
corresponding to the two sets of 320 fibers running separately to
each spectrograph.  To correct for these effects, we reject stars
outside of the $99\%$ confidence envelope resulting from the $\Delta
m$ corrections described above, and work only with half plates which
have at least 5 remaining stars (the average number is about 20).
The median $\Delta m$ offsets are calculated for each half plate
and subtracted from the values for each of the stars observed with that
half plate.

\begin{figure}
\plotone{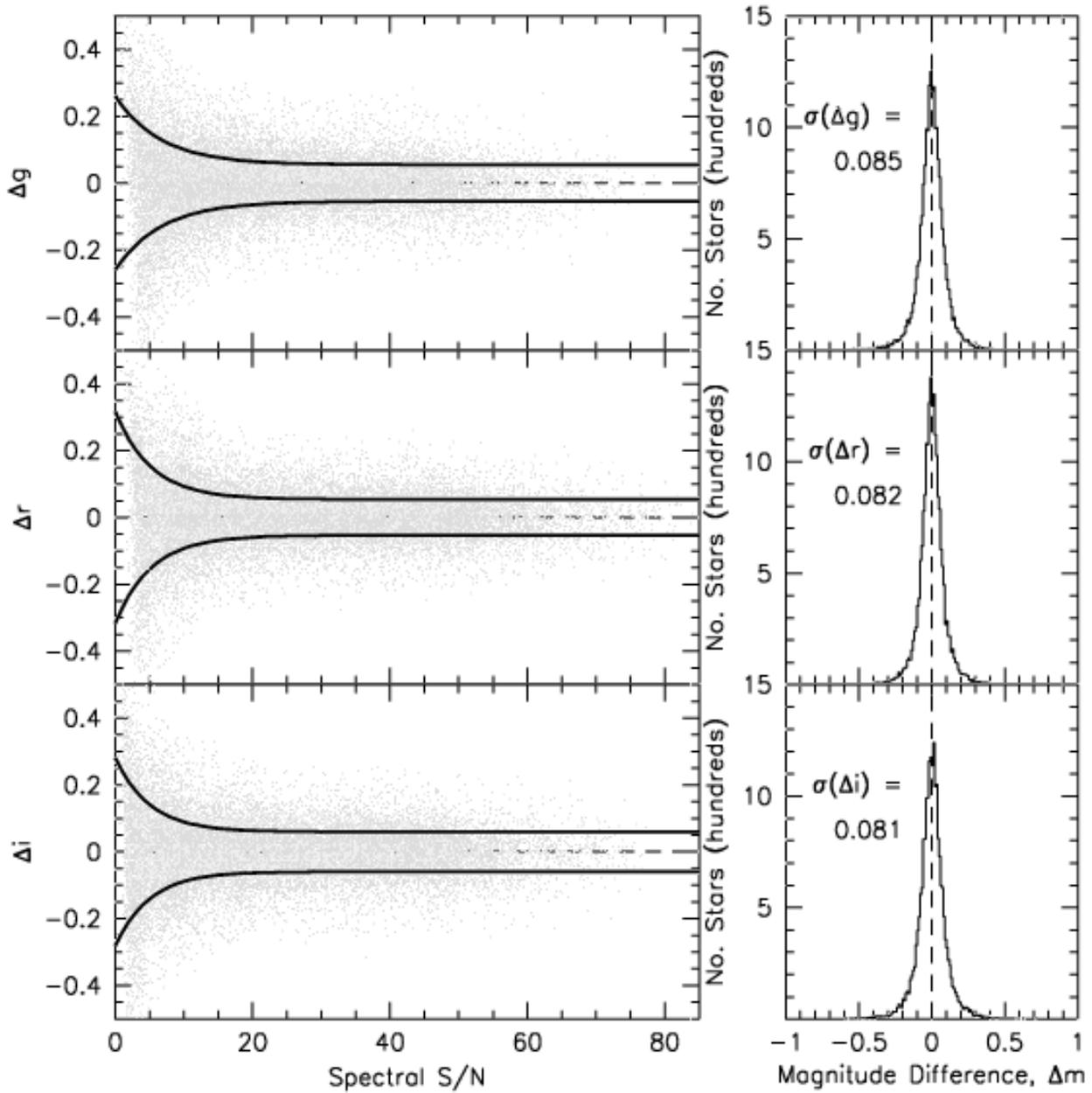}
\caption{The corrected spectroscopic minus photometric magnitudes
vs. spectral signal-to-noise ratio for stars observed on the same
spectral plates as the quasars (left).  Results for all three bandpasses
are shown.  The curves are the fits to the $68.3\%$
confidence half-width envelopes.  The right side panels show the
magnitude difference histograms for each band.  The $68.3\%$
confidence half-widths, $\sigma$ are given for a spectroscopic S/N of 10.
\label{Fig2.2}}
\end{figure}

The final corrected $\Delta m$ distributions are shown in
Fig.\,\ref{Fig2.2} as a function of spectral $S/N$.  The width of
the stellar $\Delta m$ distribution is correlated with magnitude,
but the better correlation is with spectral $S/N$.  The reason for
this is that while magnitude and $S/N$ are correlated, it is the $S/N$
which is directly related to the quality of a spectrum.  As a whole,
the $68.3\%$ confidence half-width (nominally 1 standard deviation)
is $\approx 0.08$ in each band at a spectroscopic $S/N$ of 10, which is
a substantial improvement over the initial widths of $\approx 0.13$.
Fits to the $68.3\%$ confidence half-width as a function of spectral
$S/N$ are shown in Fig.\,\ref{Fig2.2}, for which we used a function
of the form
\begin{eqnarray}
  \sigma_{S/N} = a_0 + a_1\exp(a_2 S/N)\phn,
\end{eqnarray}
where $a_0$, $a_1$, and $a_2$ are constants.  These fits are
used as statistical measurement uncertainties for the quasars
(see \S\,\ref{measure}).

\begin{figure}
\plotone{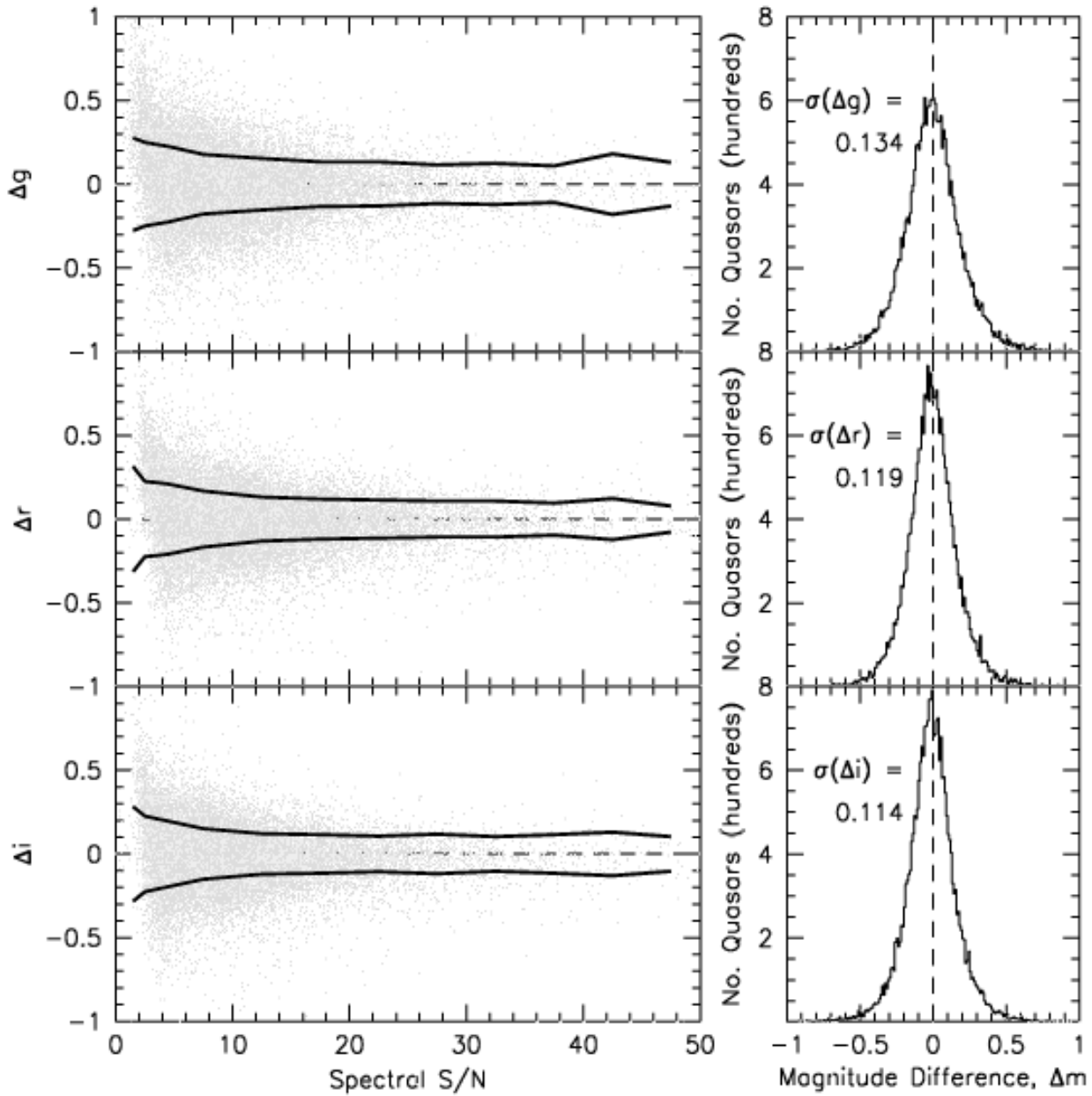}
\caption{The corrected spectroscopic minus photometric magnitudes
vs. spectral signal-to-noise ratio for the quasars (left).
Results for all three bandpasses are shown.  The curves are the
binned $68.3\%$ confidence half-width envelopes.  The right
side panels show the magnitude difference histograms for each band.
The $68.3\%$ confidence half-widths, $\sigma$, are given for a
spectroscopic S/N of 10.
\label{Fig2.3}}
\end{figure}

The same spectrophotometric corrections applied to the stars are
also applied to the quasars.  The resulting distribution of quasar
magnitude differences as a function of spectral $S/N$, and the
histograms of magnitude differences are shown in Fig.\,\ref{Fig2.3}.
The mean corrected magnitude differences for the quasars are 0.002,
-0.004, and -0.011 for the $g$, $r$, and $i$ bands respectively.
These values are small compared with the measurement uncertainties
derived from the stars.  It is possible that small differences
in the spectral energy distributions (SEDs) of stars and quasars
affect the recalibration of the quasar photometry.  However,
any effect is expected to be quite small since the same filter
transmission curves are used for both the imaging and spectroscopic
photometry, and the majority of the stars used for the recalibration
were selected as quasar candidates in the color-selected survey,
which guarantees that the SEDs are very similar.  The $68.3\%$
confidence limit half-widths of the $g$, $r$, and $i$ magnitude
difference distributions are $0.134, 0.119,$ and $0.114$ respectively
at a $S/N$ of 10, substantially larger than those of the stars.
The stars and quasars were selected to be point sources, observed
simultaneously with the same instrument, and often were selected
with the same algorithm.  The larger magnitude differences among the
quasars, therefore, demonstrate the variable nature of the quasars
in the sample.  The following sections quantify the variability
and its dependence upon many quasar parameters.

%%%%%%%%%%%%%%%%%%%%%%%%%
% Measuring Variability %
%%%%%%%%%%%%%%%%%%%%%%%%%

\section{Ensemble Variability Measurement -- The Structure
  Function\label{measure}}

The magnitude difference histograms from the previous section show
that the quasars are significantly more variable as a class than
the stars.  Assuming no stellar variability, we can use the distribution
of the stellar magnitude differences to quantify the statistical
measurement uncertainties.  Removing the width of the
stellar magnitude difference distribution in quadrature, the average
quasar magnitude differences (at a spectral S/N of 10) due to
variability in the sample are $0.103$, $0.086$, and $0.080$ in
the $g$, $r$, and $i$ bands respectively.  Measurement uncertainties
must be taken into account because they are comparable to the
values of the variability itself.  The large sizes of the samples
(both the quasars and the comparison stars) allow the measurement
uncertainty to be effectively removed.

\begin{figure}
\plotone{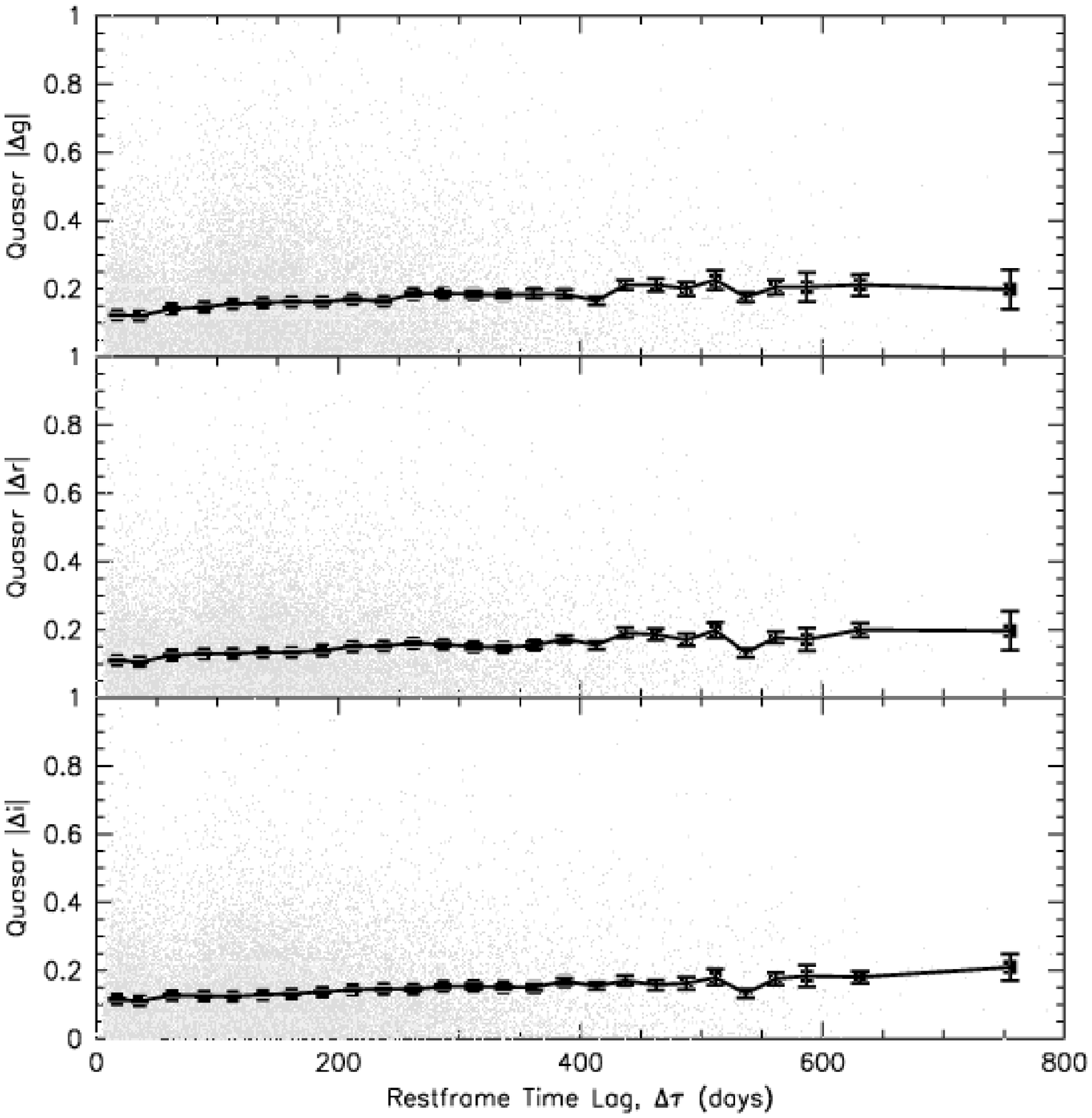}
\caption{Magnitude difference (uncorrected for measurement uncertainties)
vs.\ rest frame time delay in each of the three photometric pass bands.
The binned points show the mean values while the error bars show the 
root-mean-square deviations divided by the square root of the number
of objects in a bin.
\label{Fig3.1}}
\end{figure}

\begin{figure}
\plotone{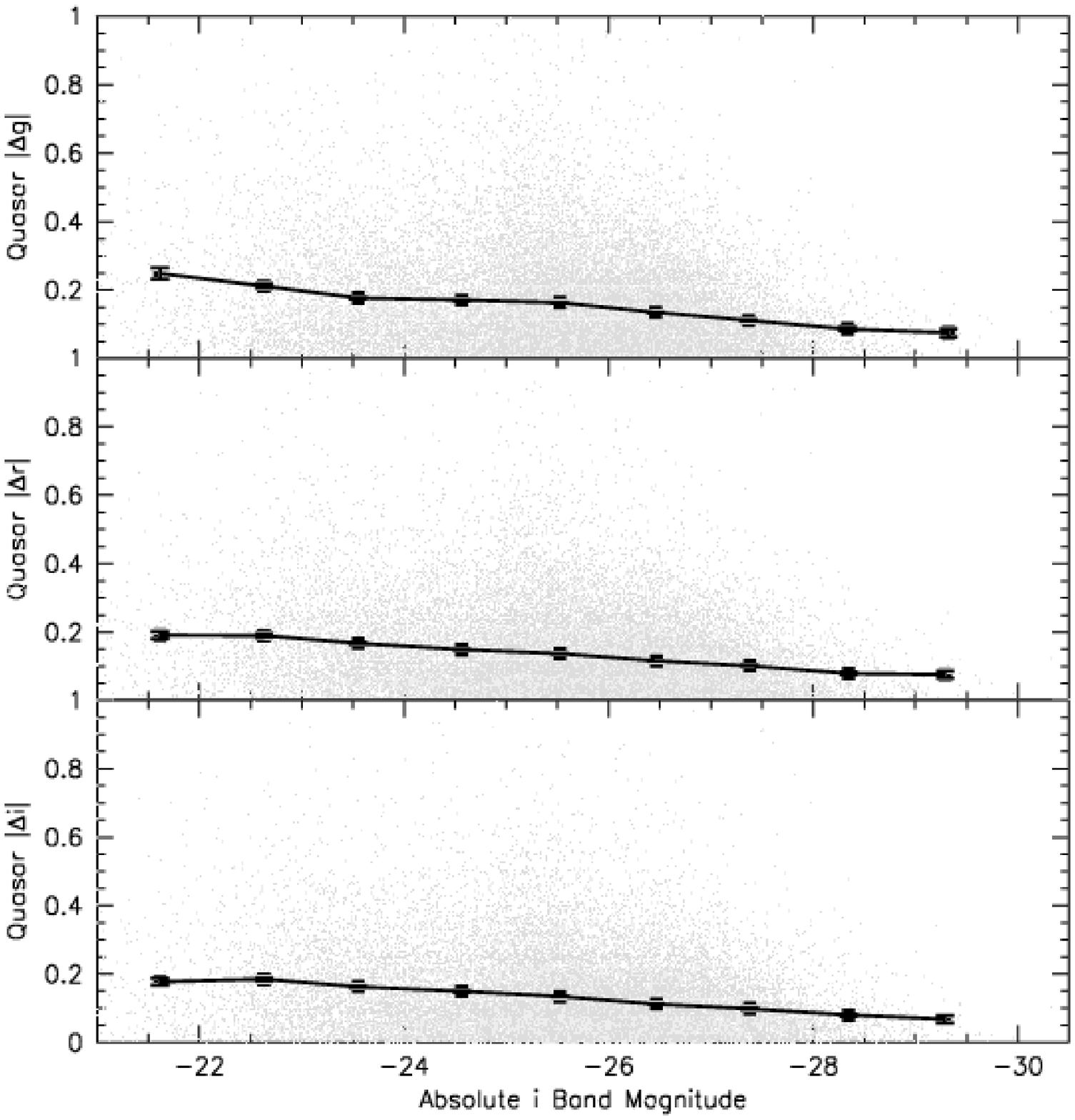}
\caption{Magnitude difference (uncorrected for measurement
uncertainties) vs.\ absolute $i$ band magnitude in each of the
three photometric pass bands.  The binned points show the mean
values while the error bars show the root-mean-square deviations
divided by the square root of the number of objects in a bin.
\label{Fig3.2}}
\end{figure}

\begin{figure}
\plotone{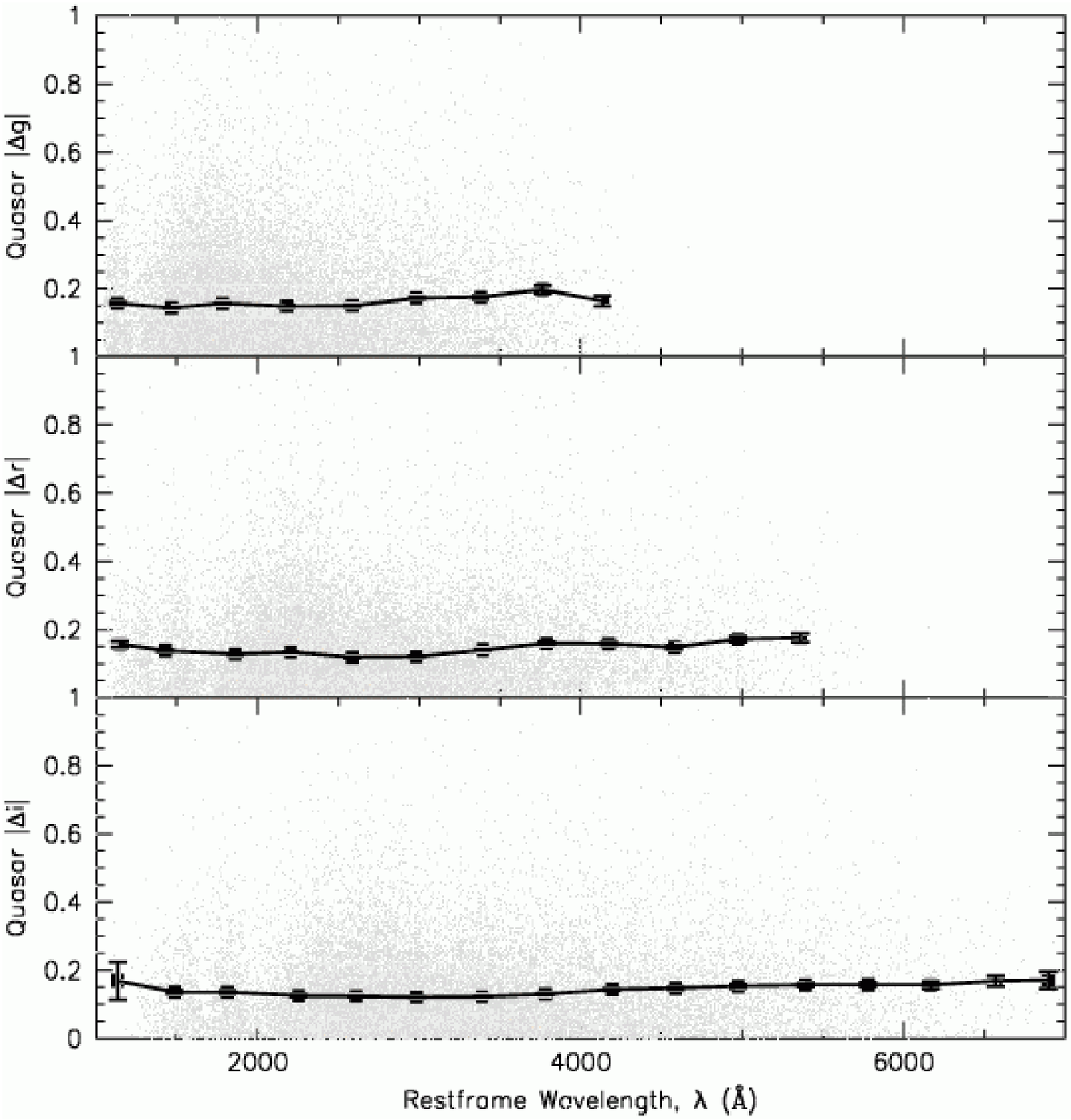}
\caption{Magnitude difference (uncorrected for measurement
uncertainties) vs.\ rest wavelength in each of the three photometric
pass bands.  The binned points show the mean values while the error
bars show the root-mean-square deviations divided by the square root of the
number of objects in a bin.
\label{Fig3.3}}
\end{figure}

\begin{figure}
\plotone{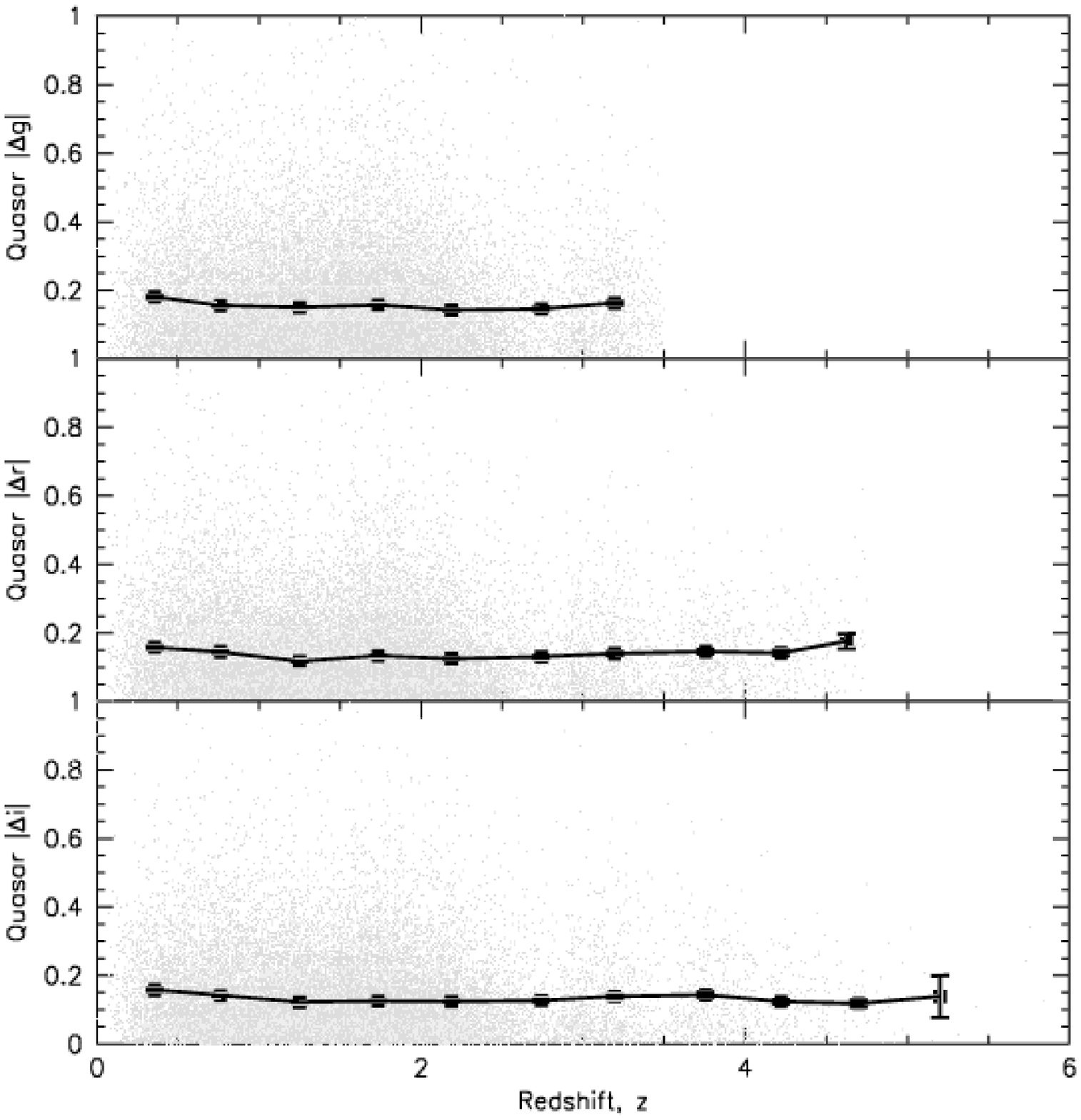}
\caption{Magnitude difference (uncorrected for measurement
uncertainties) vs.\ redshift in each of the three photometric
pass bands.  The binned points show the mean values while the
error bars show the root-mean-square deviations divided by the
square root of the number of objects in a bin.
\label{Fig3.4}}
\end{figure}

We first show the absolute values of the measured quasar
magnitude variations, uncorrected for measurement uncertainty, in
Figures\,\ref{Fig3.1} through \ref{Fig3.4}, as a function of quasar
rest frame time lag (Fig.\,\ref{Fig3.1}), absolute magnitude in
the rest frame $i$ band (Fig.\,\ref{Fig3.2}), rest frame wavelength
(Fig.\,\ref{Fig3.3}), and redshift (Fig.\,\ref{Fig3.4}).  Data in
each of the three photometric bands are shown separately.  Average
values in a set of bins are also shown.  Because flux densities in
the Ly\,$\alpha$ forest region are not representative of the true
quasar flux, we have omitted data in each band at redshifts beyond
which the Ly\,$\alpha$ forest covers the band: $z = 2.5$, 4.75,
and 6.0 for the $g$, $r$, and $i$ bands respectively.  The number of
measurements rejected for each band due to the Ly\,$\alpha$ forest
are 742, 45, and 0 for the $g$, $r$, and $i$ band respectively.
The figures show that there are several apparent correlations even
in the uncorrected data.  In particular, the average magnitude
difference increases with time lag in all three bands, and the
magnitude difference decreases with more negative absolute magnitude
(decreases with luminosity).  No trends are apparent at this stage
between magnitude difference and rest wavelength or redshift.  Again,
it is important to account for measurement uncertainty before making
any claims about the dependence of variability on any parameter.

The definition of variability used here is a statistical measure
of the magnitude difference, taking into account measurement
uncertainty.  The first application is to the dependence of
variability on rest frame time lag -- the so-called structure function.
Historically, the structure function has been the primary measure
of variability for studies of both individual quasars and quasar
ensembles.  For individual quasars with multiple sampling epochs,
the structure function is comprised of the values of the magnitude
differences for each pair of time lags in the data set, and
it is closely related to the autocorrelation function
\citep[e.g.][]{simonetti85}.
In the ensemble case, here with only two sampling epochs, the
structure function is simply the average value of the magnitude
difference for all objects with the same (or nearly the same)
time lags.  The error analysis is simpler in the ensemble case
since all of the data points are independent.

We define the ensemble variability, $V$, of a set of quasars as
\begin{eqnarray}
  V = \left(\frac{\pi}{2} {\langle \vert \Delta m \vert \rangle}^{2}
      - \langle \sigma^{2}_{S/N} \rangle\right)^{\frac{1}{2}}\phn,
\end{eqnarray}
where $\Delta m$ is the measured magnitude difference, $\sigma_{S/N}$
is the statistical measurement uncertainty of $\Delta m$ (as
a function of spectral S/N) derived from the fits to the star
measurements in \S\ref{calib}, and the brackets denote average
quantities.  The average absolute value of the magnitude difference,
along with the scaling factor of $\pi / 2$, is more robust against
the presence of outliers in the data than the average of the square
of the differences.  The values $V$ as a function of rest frame time
lag $\Delta \tau$, define the structure function, $V(\Delta \tau)$.
The same relation has been  used for the structure function in
previous variability studies \citep[e.g.][]{diclemente96}.

\begin{figure}
\plotone{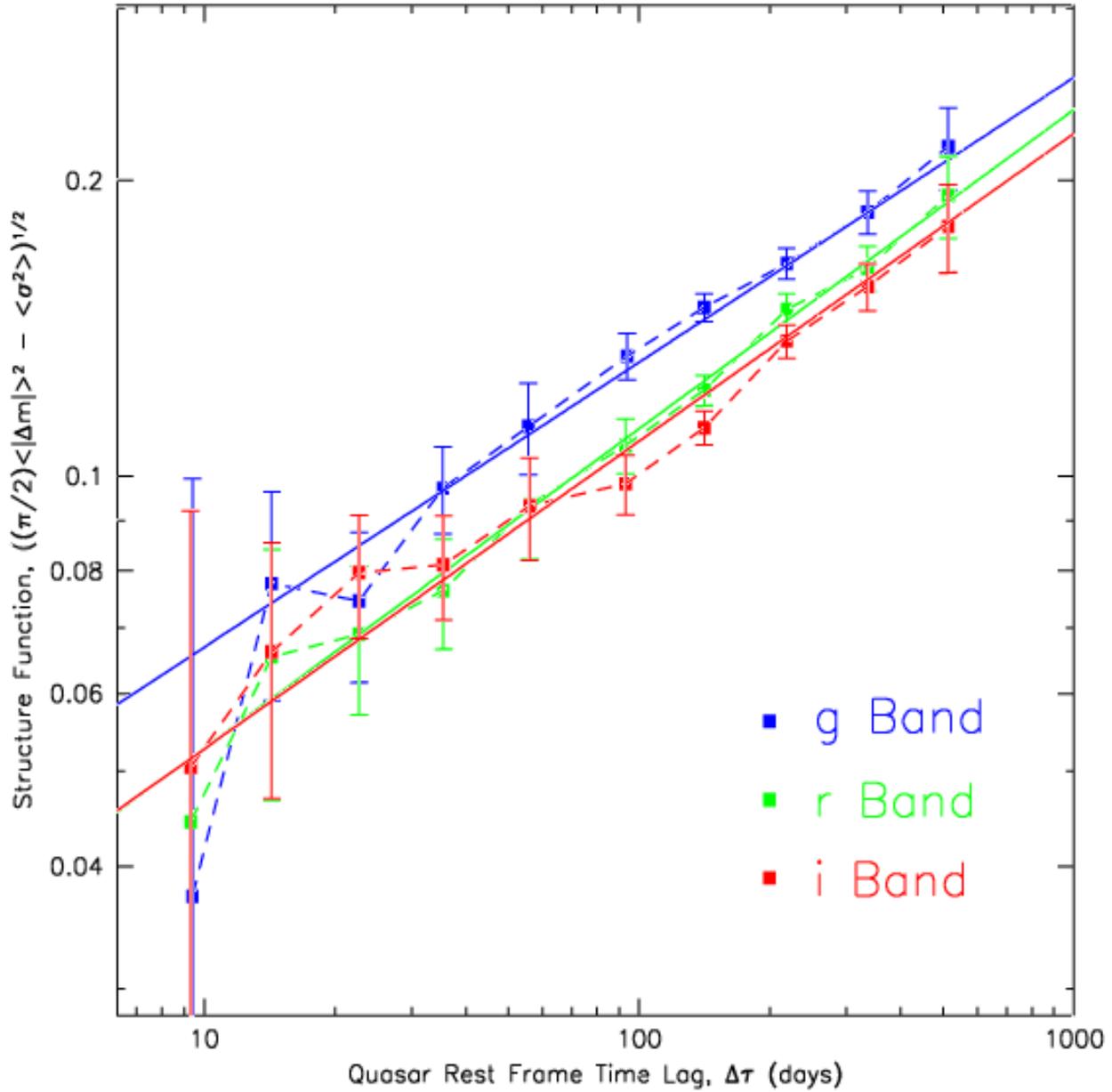}
\caption{Quasar structure functions for each of the three pass bands,
color coded by band.  No accounting has been made for any other
variability dependencies, such as luminosity, wavelength, or redshift.
Single power law fits to the data are also shown.
\label{Fig3.5}}
\end{figure}

The binned structure function for all of the quasars in the
sample for each of the three photometric bands is shown in
Fig.\,\ref{Fig3.5} with logarithmic axes.  The bins were chosen to
have equal intervals in logarithmic rest frame time lag, and to
have reasonably large numbers of objects.  The number of objects
per bin range from 241 for the shortest time lag bin covering 7
to 11 days in the $g$ band, to 7919 for the $i$ band bin covering
time lags from 111 to 176 days.  Quasars with magnitude differences
larger than $0.75$ -- just over 5 times the $1 \sigma$ width of
the distribution for quasars -- were rejected from the analysis in
order to remove outliers.  This step removes about $1\%$ of the
quasars, which is more than would be expected for a truly normal
distribution.  In a related paper \citep{ivezic03}, the distribution
of ultraviolet-excess quasar magnitude differences, including very
large differences, is discussed in more detail.  The apparently
highly variable quasars may be optically violent variables and are
valuable for follow-up studies, but the focus here is on ``typical''
quasar variability.  The error bars were determined by propagating
the root-mean-square errors $\sigma^{2}$, in the average magnitude
difference and measurement uncertainty in quadrature
\begin{eqnarray}
  \sigma(V) = \frac{1}{2}V^{-1}\sqrt{
    \pi^{2} {\langle \vert \Delta m \vert \rangle}^{2}
    \sigma^{2}(\langle \vert \Delta m \vert \rangle)
  + \sigma^{2}(\langle \sigma^{2}_{S/N} \rangle)
  }\phn.
\end{eqnarray}
Two trends are obvious: first the structure function
increases as a function of time lag -- the magnitude differences
are greater the longer the time between measurements.   Second, the
amplitude of variability is greater in the $g$ (bluest) band than in
the others, and the $r$ band amplitude is generally greater than in
the $i$ (reddest) band.  This is the variability anticorrelation with
wavelength found in a several previous studies.  We will quantify the
wavelength dependence explicitly in \S\,\ref{wavelength}, accounting
for the dependence on other parameters.  For the purposes of the
remainder of this section, the clear wavelength dependence means
that the analysis of the structure function will address the three
bands individually.

The correlation of variability with time lag has been found in numerous
previous studies, however the form of the structure function has
remained a topic of debate.  We fit the binned structure functions
with the two most common parameterizations.  The first is a power law
\citep[e.g.][]{hook94} 
\begin{eqnarray}
  V(\Delta \tau) = \left(\frac{\Delta\tau}{\Delta\tau_{0}}\right)^{\gamma}\phn,
\end{eqnarray}
where $\Delta\tau_{0}$ and $\gamma$ are constant parameters to be
determined.  This will appear as a straight line in a log-log plot
such as Fig.\,\ref{Fig3.5}.  The second is an asymptotic function
-- a constant minus an exponential -- which is the most common
parameterization of the structure function \citep[e.g.][]{bonoli79,
trevese94, hook94, enya02}
\begin{eqnarray}
  V(\Delta\tau) = V_{0}\left( 1 - e^{\frac{-\Delta\tau}{\Delta\tau_{0}}}\right)
  \phn,
\end{eqnarray}
where again $V_{0}$ and $\Delta\tau_{0}$ are to be determined.
The ``time scales'', $\Delta\tau_{0}$, whatever their values, are
simply parameters of the functions to be fit to the data, and can
not necessarily be directly compared with physical characteristic
time scales, such as those associated with accretion disks, star
bursts, or gravitational lens dynamics.

Parameter values, uncertainties, and $\chi^{2}$ values for each
of the functions in each of the photometric bands are given
in Table\,\ref{tab3.1}.  Based on the $\chi^{2}$ values, the
functional form that best fits the structure function in each band
is a power law.  The power law slopes in each of the three bands --
$0.293\pm0.030, 0.336\pm0.033,$ and $0.303\pm 0.035$ for $g$, $r$,
and $i$ respectively -- are consistent with each other within one
standard deviation of the difference.  The power law scale factors
(where the structure functions would have a value of one) are not
well constrained, mainly because the observed time lag only extends
to about 700 days.  The shape of the structure function at much
longer time lags is sometimes observed to ``flatten'' somewhat
(\citet{cristiani97, hook94, trevese94}; see \citet{hawkins02}
for a counter-example), but the data at long time lags do not
yet favor one parameterization over another.  In any case, what
we can say from this study is that a 2-parameter power law is
a good description of the data -- and a better description than
a 2-parameter exponential -- up to time lags of about 2 years.
As the SDSS proceeds, the range of time lags will eventually reach
up to about 5 years.  The rest frame time sampling will continue
to improve on all scales, and the power law form can be even more
stringently tested on longer time scales.

The wavelength dependence of the structure function becomes clearer
at longer time lags (and should become even clearer by survey end),
as would be expected from either a power law or exponential fit.
The distributions of quasar properties -- e.g.  luminosity and rest
frame wavelength -- vary with rest frame time lag because of survey
selection effects and the dependence of these properties on redshift.
In the next section we disentangle the dependence of variability
on four primary quasar parameters.

%%%%%%%%%%%%%%%%%%%%%%%%
% Primary dependencies %
%%%%%%%%%%%%%%%%%%%%%%%%

\section{Variability Dependence on Time Lag, Luminosity, Wavelength,
and Redshift\label{primary}}

\subsection{Selection Function\label{selection}}
The structure function calculated in the previous section describes
the variability of the full data set with respect to time lag.
However, variability is almost certainly also a function of quasar
luminosity and rest frame wavelength, and possibly redshift.
In order to separate the dependence of variability on multiple
parameters, the selection biases must be taken into account.
Even the structure function may not give the true dependence of
variability on rest frame time lag.  The set of quasars within a
narrow range of time lags will be populated with objects with wide
ranges of the other parameters.  In addition, since rest frame time
lag is dependent upon redshift, as are the other parameters, the
distributions of quasar parameter values will be correlated with
the time lag.  For example, high redshift quasars will generally
have shorter time lags in the rest frame than lower redshift quasars.

\begin{figure}
\plotone{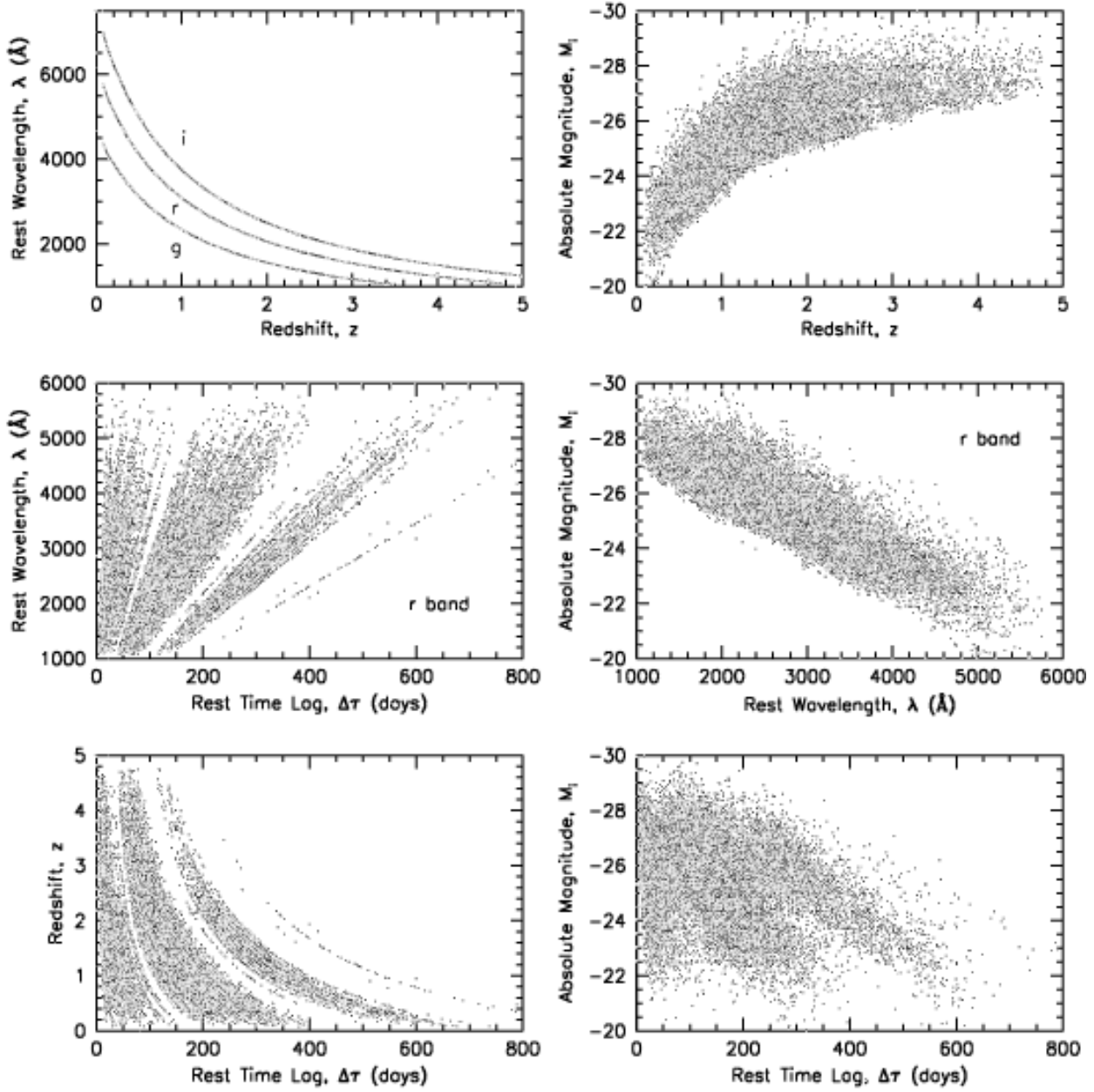}
\caption{Projected parameter values for all of the quasars.  This
effectively shows the selection function in the parameter space
given by rest frame time lag, $\Delta\tau$, redshift, $z$, absolute
$i$-band magnitude, $M_{i}$, and rest wavelength, $\lambda_{rest}$.
The passbands are indicated when necessary.  For clarity, only
the $r$ band data are shown for the middle two plots.
\label{Fig4.1}}
\end{figure}

The selection function -- that is, the region of parameter space
occupied by the data set -- is shown in projected planes in
Fig.\,\ref{Fig4.1}.  A number of artificial correlations are
evident and are due both to the survey selection criteria, and to the
dependence of luminosity, rest frame time lag, and rest wavelength on
redshift.  Variability information can obviously only be obtained in the
regions of the four dimensional parameter space containing a
statistically sufficient number of objects.

In order to determine how variability is related to a single parameter,
the space was divided into small regions in three dimensions, and
the variability calculated as a function of the remaining parameter
in each of the slices.  The condition that there be enough quasars to
reliably measure variability was the primary limiting factor for the
bin sizes.  For each parameter there is then a set of variability
relations, each set representing the results of restricting the
ranges of the other 3-parameters.  As shown in the remainder
of this section, in most cases variability trends are clear even in
independent restricted data sets.

If we make the assumption that the equations describing the
multi-parameter dependence of variability are separable, the 
results from each of the slices may be scaled in the single 
parameter ranges where they overlap, in order to find the 
variability dependence upon a single parameter.  That is, the
form of the variability dependence is assumed to be
\begin{eqnarray}
  V(\Delta\tau,M_{i},\lambda,z) = v(\Delta\tau) \times v(M_{i})
  \times v(\lambda) \times v(z)\phn,
  \label{separable}
\end{eqnarray}
where $V$ is calculated as in the previous section.  While this form
is not necessarily correct, it greatly simplifies the analysis, and
the relatively simple relations found for each parameter suggest that
it is not far from reality.  In the following subsections, we show
the unscaled variability relations for single parameters in the
independent restricted data sets, then show the results after scaling
the independent sets together assuming equation \ref{separable}.
The scaled relations are fit with relatively
simple descriptive functions for each parameter.

%
% Time Lag
%
\subsection{Time Lag\label{timelag}}
\begin{figure}
\plotone{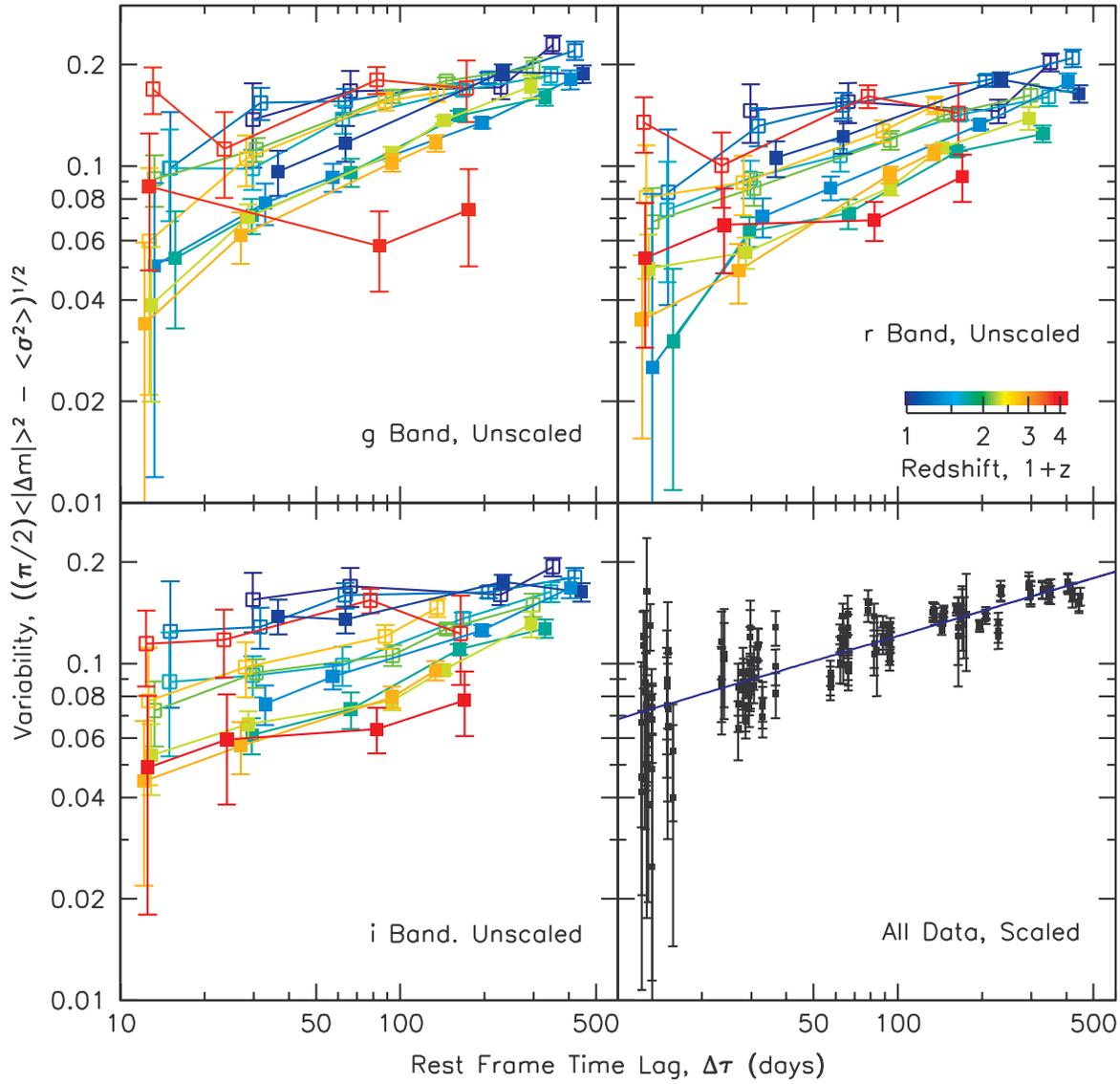}
\caption{Variability as a function of rest frame time lag for
independent slices of data (upper left, upper right, and lower
left panels).  Colors indicate redshift, with redder colors showing
data from higher redshift ranges, according to the color key.
Solid and open points show results from slices with higher and
lower luminosities respectively, within the same redshift range.
Results from the three photometric bands are given separately,
which effectively restricts the rest wavelength ranges to small
values in each slice.  The lower right panel shows all of the scaled
data points, along with the best fit power law for variability
vs. $\Delta\tau$.
\label{Fig4.2}}
\end{figure}

We focus first on the dependence of variability on rest frame
time lag, independent of the other 3 parameters.  The full quasar
sample was first separated into 6 redshift bins, each with an equal
spacing in logarithmic $1+z$.  The redshift bin sides are: 0.185,
0.499, 0.895, 1.395, 2.028, 2.829, and 3.840.  The quasar sample
in each redshift bin was then divided into two halves separated at
the median absolute magnitude of the quasars in the redshift bin:
$M_{i,median} =$ -22.96, -24.07, -25.24, -26.07, -26.73, and -27.09.
Taking each photometric band separately for the quasars in each of
the 12 redshift/absolute magnitude bins restricted the quasar rest
wavelengths to small ranges.  The procedure produced 36 independent
data sets confined to small ranges of redshift, absolute magnitude,
and rest wavelength, but unrestricted with respect to rest frame
time lag.  The variability amplitude and uncertainty as a function of
time lag were determined as in \S\,\ref{measure}, for the quasars in
each of the 36 data sets independently.  The time lag bins were set
to have a constant logarithmic time width, as in \S\,\ref{measure},
but with twice the width to accommodate the smaller number of
objects per bin.  The rest frame time lag bin sides, in days, are:
7.0, 17.6, 44.3, 111.1, 278.7, and 699.2.  Unphysical (imaginary)
values of the variability amplitude occurred in a small number
(3) of cases in which the number of quasars was relatively small.
In most cases when this occurs the number of quasars is 5 or fewer.
For all further analysis, binned data sets containing fewer than
10 quasars or which produce imaginary values of the variability
amplitude are rejected.  For each of the 36 data sets,
the variability with respect to rest frame time lag is an
independent structure function over which the absolute magnitude,
rest wavelength, and redshift do not vary greatly.  The results
are shown in Fig.\,\ref{Fig4.2}.  Results for the three photometric
bands, corresponding to restricted ranges of rest wavelength, are
shown in separate panels.  The average redshift of the quasars
contributing to each structure function is indicated by color,
with redder colors representing progressively higher redshifts.
The structure functions containing the more luminous halves of the
quasar data sets are shown with solid points, and the less luminous
halves with open points.

It is clear from Fig.\,\ref{Fig4.2} that variability is an increasing
function of rest frame time lag at all redshifts, absolute magnitudes,
and rest wavelengths.  Two other trends can also be seen: the less luminous
quasars vary more than their more luminous counterparts (nearly all of the
open points lie above the solid points within the same redshift bin);
and quasars vary more at shorter wavelengths, confirming what was shown by the
unrestricted structure functions in \S\,\ref{measure}.

Under the assumption that the variability as a function of time lag
is separable from the other dependencies, the individual structure
functions can be scaled together in the time lag regions where
they overlap.  Using as a reference a structure function near the
middle of the redshift, luminosity, and wavelength distributions,
all 36 structure functions (excluding bins with too few objects)
were scaled such that the sum of the products of the amplitudes and
the time lag bin widths (the areas under the curves) were equal.
The scaled points are shown in the last panel of Fig.\,\ref{Fig4.2}
along with the best fit power law.  The parameter fits to a power
law and exponential are given in Table\,\ref{tab4.1}.  A 2-parameter
power law provides a very good fit and is better than the asymptotic
(exponential) form. The power law fit has a slope of $\gamma =
0.246\pm0.008$, which is comparable to but shallower than the values
found for the unrestricted structure functions in \S\,\ref{measure}.
Scaling the data points will change the characteristic time scale of
the function (the time lag at which the power law amplitude would be
unity), but not the power law slope.  That the slope is relatively
close to those found for the unrestricted structure functions is
due to the offsetting variability dependencies on luminosity and
rest wavelength.  From Fig.\,\ref{Fig4.1}, longer time lags are
statistically populated by more luminous objects, which vary less,
but at shorter wavelengths, where the variability is greater.
The significance of the power law slope in relation to variability
models is discussed in \S\,\ref{discussion}.

\subsection{Absolute Magnitude\label{absmag}}
\begin{figure}
\plotone{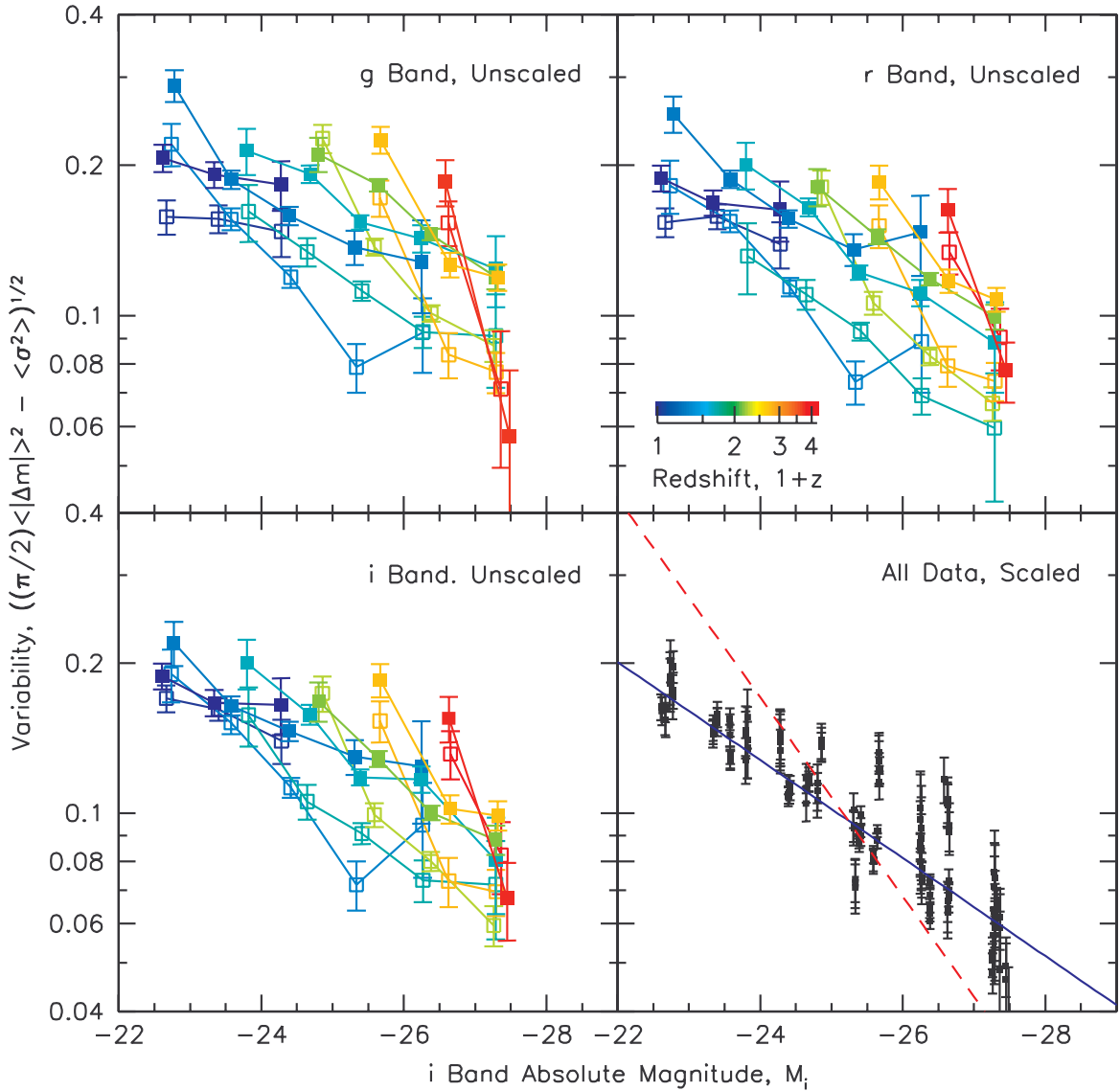}
\caption{Variability as a function of absolute magnitude for
independent slices of data (upper left, upper right, and lower left
panels).  Colors indicate redshift, with redder colors showing data
from higher redshift ranges, according to the color key.  Solid and
open points show results from slices with longer and shorter time
lags respectively, within the same redshift range.  Results from
the three photometric bands are given separately, which effectively
restricts the rest wavelength ranges to small values in each slice.
The lower right panel shows all of the scaled data points, along with
a best fit generalized Poissonian function (solid) and Poissonian
function with the power law index fixed at $1/2$ (dashed).
\label{Fig4.3}}
\end{figure}

The luminosity (absolute magnitude) dependence of variability is
separated from the other parameters in a similar way to the time lag
in the previous subsection.  The full quasar sample was separated into
the same 6 redshift bins, and each separate quasar sample was further
divided into two halves at the median rest frame time lag of the
quasars in the redshift bin.  The median rest frame time lags, in days,
for each redshift bin are: 249.9, 192.0, 157.1, 122.9, 100.5,
and 73.6.  Again considering the three photometric
bands separately restricted the rest wavelengths of each data set
to small ranges.  This produced 36 independent quasar subsamples
unrestricted with respect to absolute magnitude.  The variability
amplitude as a function of absolute magnitude, in bins one magnitude
in width, is shown for each quasar
subsample in Fig.\,\ref{Fig4.3}.  The variability amplitude is an
increasing function of absolute magnitude (brighter objects vary less)
for nearly every subsample.  Also seen in Fig.\,\ref{Fig4.3} are the
time lag dependence -- variability amplitudes are greater at longer
time lags (filled symbols) than short time lags (open symbols) -- and
a wavelength dependence, seen most easily in the amplitude differences
between the $g$ and $i$ bands.  This is a clear demonstration that
the well-known luminosity-variability anticorrelation is not simply
due to time lag or rest wavelength selection effects.

% Is there also a redshift dependence here -- the higher redshift
% bins seem to have a steeper slope.  Higher-z ==> shorter dtau,
% but steeper slope means a greater scale factor, which can only be
% provided by the redshift.

The data sets were again scaled such that that the areas under the
curves were equal.  In each redshift bin, the six individual sets (3
wavelength and 2 time lag bins) were scaled to have identical sums
of the product of the absolute magnitude bin width and variability
amplitude.  Then proceeding from the lowest to highest redshifts,
the scaled redshift sets were rescaled to the adjacent redshift set
so that the areas under the curves were equal in regions where the
absolute magnitude coverage overlapped.  The scaled data points
are shown in the last panel of Fig.\,\ref{Fig4.3}.  A straight
line can be fit to the data points, but such a description is
unphysical since at large luminosities (large negative absolute
magnitudes) the function becomes negative.  In so called Poissonian
or discrete-events models the relative luminosity variability,
$\delta L/L$ is expected to vary with luminosity as $\delta
L/L \propto L^{-\beta}$, where $\beta = \frac{1}{2}$ in general
\citep[e.g.][]{cid00}.  This relationship translates into
the absolute magnitude form
\begin{eqnarray}
  v(M_{i}) \propto 10^{\beta M_{i}/2.5}\phn.
\end{eqnarray}
%
%where $M_{0}$ corresponds to the scale factor in the differential
%luminosity equation.
This function and one in which $\beta$ is held fixed at 0.5
were fit to the data.  Both fits are shown in Fig.\,\ref{Fig4.3}.
The logarithmic equation, with a best fit of $\beta = 0.246\pm0.005$,
fits the data as well as a straight line, and avoids the problem of
negative values.  The Poissonian prediction of $\beta=0.5$ gives a
poor fit and is clearly inconsistent with the data.  Scaling the
individual data sets, which accounts for arbitrary contributions
from variability dependencies on time lag, rest wavelength, and
redshift, does not change the value of $\beta$.  We will discuss
the implications for Poissonian models further in \S\ref{discussion}.

\subsection{Rest Wavelength\label{wavelength}}
\begin{figure}
\plotone{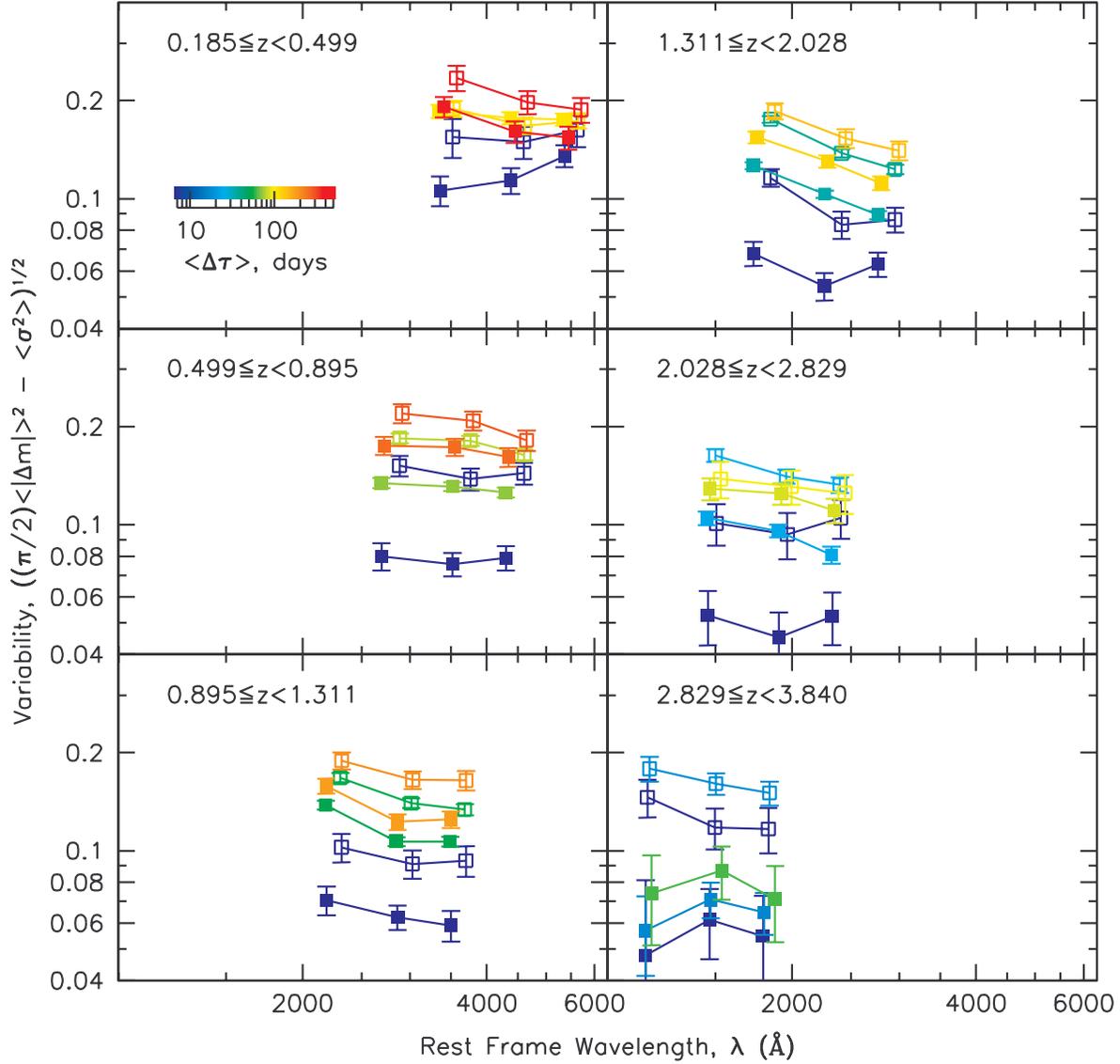}
\caption{Variability as a function of rest frame wavelength for
independent slices of data.  Colors indicate average rest frame time
lag, with redder colors showing longer time lags, according to the
color key.  Solid and open points show results from slices with
more luminous and less luminous quasars respectively, within the
same redshift and time lag ranges.  Results from the six redshift
slices are shown in separate panels.
\label{Fig4.4}}
\end{figure}

The rest wavelength dependence of variability was isolated for
subsamples selected to cover small ranges in redshift, time lag, and
absolute magnitude.  The redshift and absolute magnitude bins are the
same ones used to isolate the time lag dependence (12 separate bins,
see \S\ref{timelag}).
The data in each of these bins were divided into three separate samples
in the plane of absolute magnitude and time lag, according to the following
lines: $M_{i} = 0.085 {\Delta\tau} - 31.25$ and
$M_{i} = 0.024 {\Delta\tau} - 31.42$.
%%
%%lag bins,
%%the ranges of which depend upon the time lags covered by
%%the data. 
%%
This gives 36 independent data sets covering small ranges of redshift,
absolute magnitude, and time lag.  All but one of the data sets
contain enough objects to compute reliable variability measurements.
Each data set samples three separate rest wavelength points given
by the effective rest wavelengths of the three photometric bands.
The variability amplitude as a function of the rest frame wavelength
for each set is shown in Fig.\,\ref{Fig4.4}.  The average rest frame
time lag in each set is color coded (longer time lags are redder),
the more luminous half of a redshift/time lag bin is shown with solid
points, and each redshift subset is shown in a separate panel.

In most cases, the variability amplitude decreases with wavelength
as expected from previous analysis.  The cases in which the opposite
happens occur at short time lags and very low or very high redshifts,
but there are too few cases to make any claims about deviations from
the general trend.  The time lag and luminosity dependencies are
also evident from the figure.

\begin{figure}
\plotone{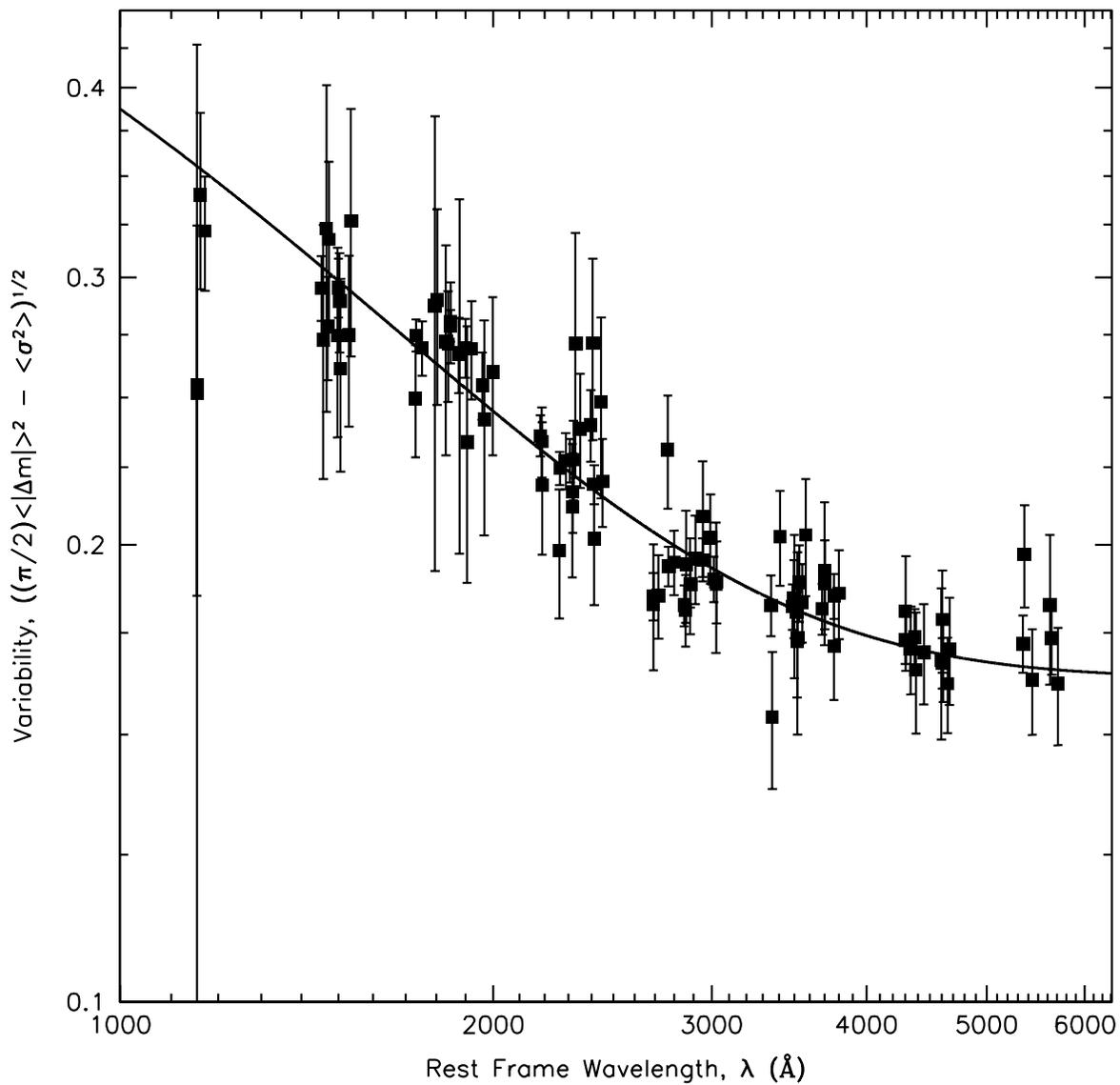}
\caption{Scaled variability amplitude points from Fig.\,\ref{Fig4.4}
as a function of rest frame wavelength.  The line shows the best fit
exponential function as described in the text.
\label{Fig4.5}}
\end{figure}

The data points were scaled in a manner similar to that in
\S\,\ref{absmag}.  For each redshift bin, the six sets of points (3
time lag bins and 2 absolute magnitude bins) were scaled to the same
area under the curves and then, moving from low to high redshift,
the points were rescaled to match the appropriate area in the adjacent
redshift bin.  In this case, since the rest wavelength bin limits are
not equal for the separate redshift bins, the three wavelength points
in each set were connected by two straight lines, and the area under
the lines was calculated in the regions where they overlapped the
wavelengths of the adjacent redshift bin points.  The scaled points
as a function of rest wavelength are shown in Fig.\,\ref{Fig4.5}.
If the variability is due to a simple change in the index of a single
power law, we would expect the wavelength dependence to be
\begin{eqnarray}
  v({\lambda}) = -2.5{\Delta\alpha_{\lambda}} \log({\lambda}) + C\phn,
\end{eqnarray}
where ${\Delta\alpha_{\lambda}}$ is the difference in the wavelength
power law index, and C is a constant related to the pivot wavelength
(where the two power laws intersect) presumed to be much longer
than the observed wavelengths.  Figure \,\ref{Fig4.5} is plotted
with a logarithmic wavelength axis, and so the relation would be
seen as a straight line.  A single straight line is an adequate fit
from the shortest wavelengths up to about 4000{\AA}, but does not
account well for the longer wavelength end.  Contamination from host
galaxy light at longer wavelengths would have the opposite effect
-- cause the variability to decrease even faster with wavelength.
A 3-parameter exponential function, although physically unmotivated,
fits the data well
\begin{eqnarray}
  v({\lambda}) = a_0\exp(-{\lambda}/{\lambda_0}) + a_1\phn,
\end{eqnarray}
with parameter values $a_0=0.616\pm0.056$, $\lambda_0=988\pm60${\AA},
and $a_1=0.164\pm0.003$.

\subsection{Redshift\label{redshift}}
\begin{figure}
\plotone{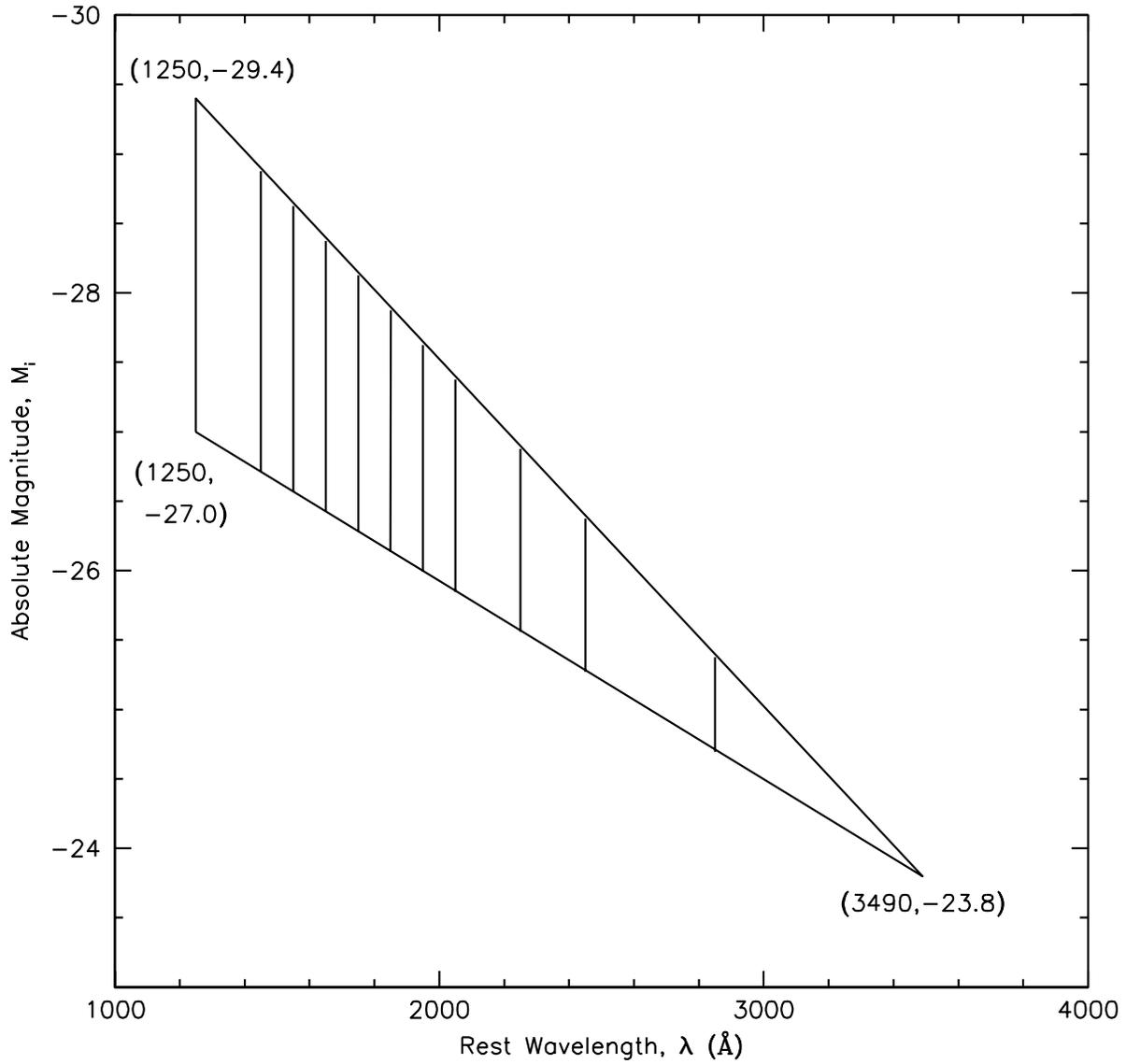}
\caption{The triangle shows the region of the wavelength-absolute magnitude
plane that is simultaneously well covered by quasar observations in all three
photometric bands.  The vertical lines -- separated by 100, 200, 400, or
800{\AA} -- show the bin sides used to separate the quasar sample by
wavelength.
\label{Fig4.6}}
\end{figure}

The redshift dependence of variability is more difficult to
isolate from the time lag, absolute magnitude, and wavelength
dependencies.  The reason for this can be seen from inspection
of Fig.\,\ref{Fig4.1}.  For example, samples restricted to a
narrow range of rest wavelengths will have 3 independent redshift
intervals (taking the three bands separately), but the corresponding
absolute magnitude ranges for the redshift intervals may not
overlap appreciably.  So to isolate the redshift dependence, we
first found a region of wavelength-absolute magnitude space which
is covered by quasar data in all three bands.  Fig.\,\ref{Fig4.6}
shows the wavelength-absolute magnitude plane and the selected
region which is bounded by a triangle with corners $(\lambda,M_{i}):
(1250,-27.0), (1250,-29.4)$, and $(3491,-23.8)$.  Outside of this
region quasar data are generally available for only 2 or fewer of
the bands.  The data contained in this region were then separated
into 16 wavelength bins, shown in Fig.\,\ref{Fig4.6}, at intervals
of 100, 200, 400, or 800{\AA}, depending on the number of objects
contained in the bin.  For each bin of wavelength separated data,
objects were selected from a single range of time lags chosen so
that the average absolute magnitudes, time lags, and wavelengths
were about equal for data in each of the three photometric bands.
Taking each of the photometric bands separately for a restricted data
set gives a wide range of redshifts, while keeping the ranges of
time lag, absolute magnitude, and rest wavelength nearly constant.

\begin{figure}
\plotone{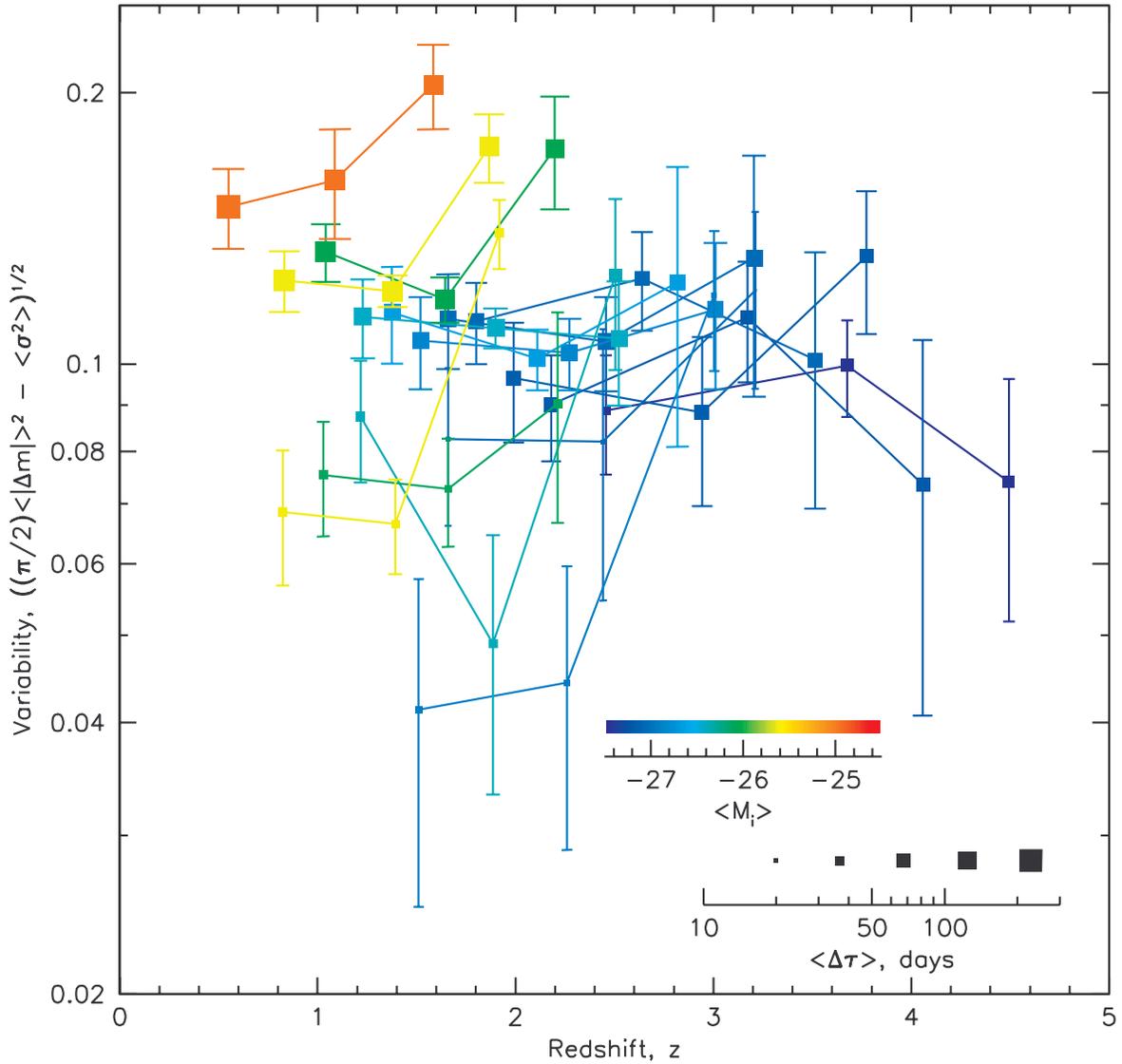}
\caption{Variability as a function of redshift for independent sets
of data.  Sets of three connected points show the variability amplitude
for data sets with similar distributions of rest wavelength, absolute
magnitude, and time lag, but different average redshifts.  Colors
correspond to the average absolute magnitude for each set, with
bluer colors indicating brighter absolute magnitudes, according to the
color key.  The point sizes correspond to average time lags,
according to the point size scale.
\label{Fig4.7}}
\end{figure}

The variability amplitudes for all 16 data sets (each with three
redshift points) are shown in Fig.\,\ref{Fig4.7}.  The number
of objects contributing to each point ranges from 65 to 1150,
with a mean number of 239.  Lines connect the points belonging to
data sets with nearly the same parameter values but at different
redshifts.  The color of the points and lines corresponds to the
average absolute magnitude of the data set, with bluer colors
representing brighter absolute magnitudes.  The point sizes
correspond to the average time lag of the objects contributing to
the points.  The absolute magnitude correlation, discussed above,
is evident in Fig.\,\ref{Fig4.7}, but it is partly counteracted by
the nearly monotonically increasing rest wavelength with average
absolute magnitude (seen from Fig.\,\ref{Fig4.6}), and the generally
longer average time lags associated with fainter average absolute
magnitudes.

\begin{figure}
\plotone{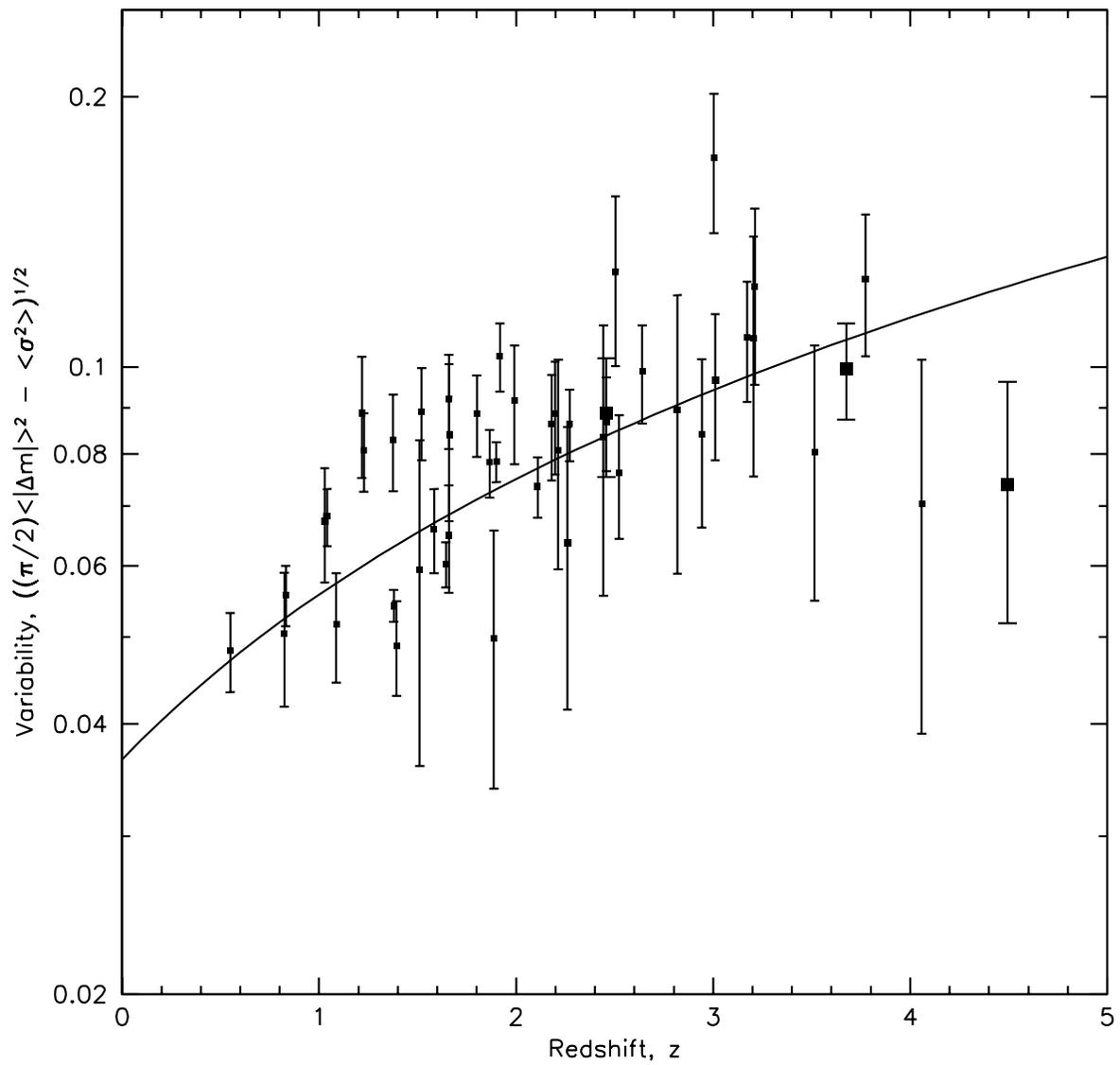}
\caption{Scaled variability amplitude as a function of redshift.
The best fit line (to linear redshift and variability amplitude)
is also shown.
\label{Fig4.8}}
\end{figure}

What is of interest here is the dependence of the variability on
redshift.  The results are fairly noisy and it is difficult to detect
any clear trend with redshift.  The sets of points were scaled by
matching the areas under the curves of adjacent data sets, as in the
previous subsections, starting with the sample with the shortest rest
wavelengths.  The scaled data points are shown in Fig.\,\ref{Fig4.8}.
There is a correlation between the scaled variability and redshift --
quasars appear to be more variable at higher redshifts.  The Spearman
rank correlation probability that the points are uncorrelated is
less than $10^{-4}$, even after accounting for the reduction of
the number degrees of freedom by the number of restricted data sets
(16).   A straight line fit to the data points (linear in redshift
and variability amplitude) gives
\begin{eqnarray}
  v(z) = (0.019\pm0.002)z + (0.037\pm0.005)\phn.
\end{eqnarray}

The correlation, although significant, is weak enough that it
could easily have gone unnoticed in previous variability studies,
especially since most of them suffer from a lack of sufficient sample
overlap needed to test the redshift relationship independently
of other parameters.  The redshift evolution of variability would
have serious consequences for a number of currently proposed models.
If the effect is intrinsic, the quasar population or the variability
mechanism is changing over time.  External causes are also possible,
such as gravitational microlensing which may increase with redshift
since more potential lenses would be available.  The variability
correlation with redshift is discussed further in \S\ref{discussion}.

%%%%%%%%%
% Color %
%%%%%%%%%

\section{Color Dependence\label{color}}

Evidence from previous ensemble studies
\citep{trevese01,giveon99,edelson90} suggests that the spectral energy
distribution of quasars becomes harder (bluer) in bright phases.
Indirect evidence also comes from the fact that there is a strong
wavelength dependence upon variability (\S\,\ref{wavelength}).
This could happen for example if the index of a power law component
of the continuum changes with luminosity \citep{trevese02}.

\begin{figure}
\plotone{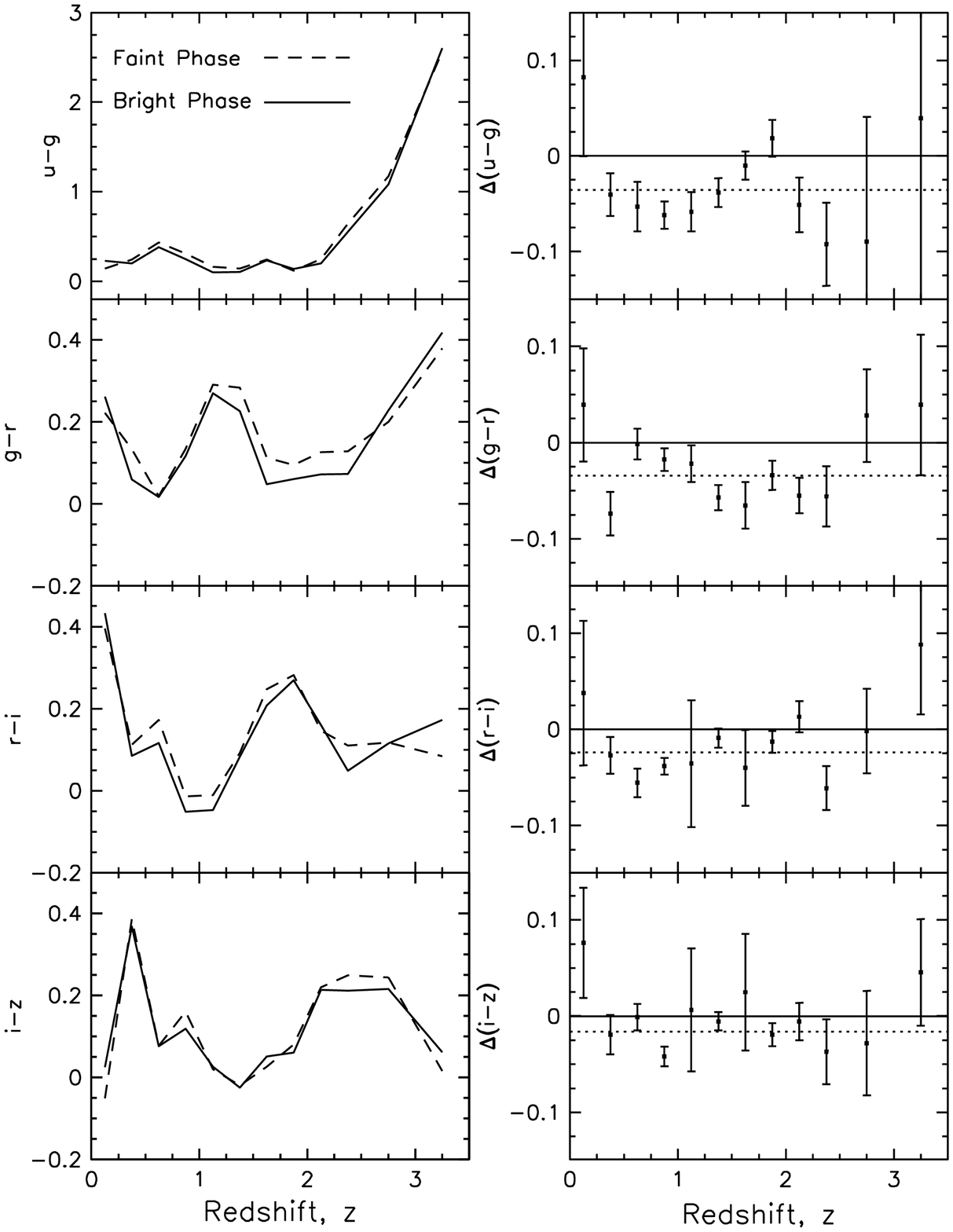}
\caption{Average faint phase vs. bright phase colors of quasars as
a function of redshift (left).  The structure in the plots is mainly
due to spectral features redshifting into and out of the various
passbands.  The plots on the right side show the color difference
between the bright and faint phases as a function of redshift,
along with the uncertainties in each bin.  A more negative value
implies a bluer spectrum in the bright phase.
\label{Fig5.2}}
\end{figure}

Quasar colors are a strong function of redshift \citep{richards01},
since various spectral features move into and out of the photometric
bandpasses with redshift changes.  A pure power law spectrum would
have a single set of colors independent of redshift.  The observed
quasar color structure is mainly due to the presence of strong
emission features, especially broad Fe\,{\sc ii} complexes, as well
as the Ly\,$\alpha$ forest.  Figure\,\ref{Fig5.2} shows the average
imaging photometric colors of quasars as a function of redshift in
two samples selected to be either brighter or fainter by at least 3
standard deviations in at least one of the $g,r,i$ bands relative
to the spectrophotometry.  We use the imaging photometric colors
rather than the spectroscopic since they are more precise, and two
more colors are available.  Although the color differences are small
($\sim 0.03$), at most redshifts up to at least $2.5$ and for each color
the bright phase sample is bluer than the faint phase sample.
Also shown in Fig.\,\ref{Fig5.2}
are the color differences of the bright phase minus the faint phase
samples as a function of redshift.  Both the binned and average color
differences are shown.  The color differences are increasingly larger
for shorter wavelength bands, i.e.\ quasars in bright phases are
bluer than those in faint phases, and they are even bluer at shorter
wavelengths.

That the color change persists at high redshift also indicates that
it cannot be accounted for solely by a non-variable red spectral
component, such as the quasar host galaxy.  Such a component would
contribute a higher fraction of the total quasar light in the faint
phases, making quasars appear redder than in the bright phases.
Any reasonable host galaxy spectral energy distribution and luminosity
would contribute very little light to the bluest bands, and would
quickly be redshifted out of the other bands.  By a redshift of 0.5,
there should be almost no significant contamination from the host
galaxies in any of the passbands.  A host galaxy component cannot
account for the wide range of redshifts over which the color difference
is significant.

%%%%%%%%%%%%%%%%%%%%%%%%%%
% Secondary Dependencies %
%%%%%%%%%%%%%%%%%%%%%%%%%%

\section{Variability of Radio, X-ray, and Broad Absorption
Line Quasars\label{secondary}}

There is evidence from previous studies that the variability
amplitude of quasars varies among different subclasses, such as those
with radio emission \citep{helfand01,eggers00,garcia99,pica83} or
broad absorption line systems \citep{sirola98}.  The entire class
of highly variable blazars, for example, is defined in part by
X-ray and radio emission \citep[e.g.][]{ulrich97}.  Here we examine
the variability of broad subclasses of quasars in comparison to
carefully selected control samples.

%
% Radio
%
\subsection{Radio Detected Quasars\label{radio}}

%TODO: check if the radio sample can be sub-divided into RL bins
%  This would require extracting the radio flux for each object.
% DONE! 12/17/02

Some SDSS quasar candidates are selected as optical matches to radio
sources in the FIRST survey \citep{white97, ivezic02}.  About $10\%$ of the
verified quasars in the sample have counterparts in the FIRST survey.
In the areas of the SDSS sample covered by the FIRST survey at the
time the quasar candidates were selected, there are 1553 verified
quasars with FIRST catalog matches.  To test whether quasars with
detected radio emission are more or less variable than those without,
we have extracted a control sample of quasars without matches in the
FIRST catalog.  The non-radio-detected control sample was constructed
to have the same redshift, luminosity, and time lag distribution
as the radio-detected sample, by matching each radio quasar with
a non-radio quasar with nearly the same redshift, magnitude, and
time lag.  The standard deviations of the redshift, magnitude, and
time lag differences are $\sigma(\Delta z) = 0.03$, $\sigma(\Delta m)
= 0.04$ mag., and $\sigma(\Delta(\Delta \tau)) = 3.6$ days, and in
no case were the differences allowed to be greater than $0.5$, $0.5$
mag., and $40$ days respectively.  Of the FIRST matched quasars,
1376, 1389, and 1388 were able to be matched with counterparts in
the non-FIRST sample in the $g$, $r$, and $i$ bands respectively.
Kolmogorov-Smirnov tests comparing the redshift, magnitude, and time
lag distributions of the radio and control samples show that they
are statistically indistinguishable.  This also guarantees that the
wavelength coverages of the samples are nearly identical.

A radio-loud (not simply radio-detected) sub-sample and its
corresponding non-radio control sample were also generated.  Radio
loudness is defined here as the ratio of the rest frame $5$GHz to
$4500${\AA} flux-densities \citep[e.g.][]{sramek80}, and a quasar is
deemed radio-loud if the ratio is at least 100.  The sample sizes of
the radio-loud quasars and the matched radio-quiet control quasars
are 492, 530, and 546 objects for the $g$, $r$, and $i$ bands respectively.

\begin{figure}
\plotone{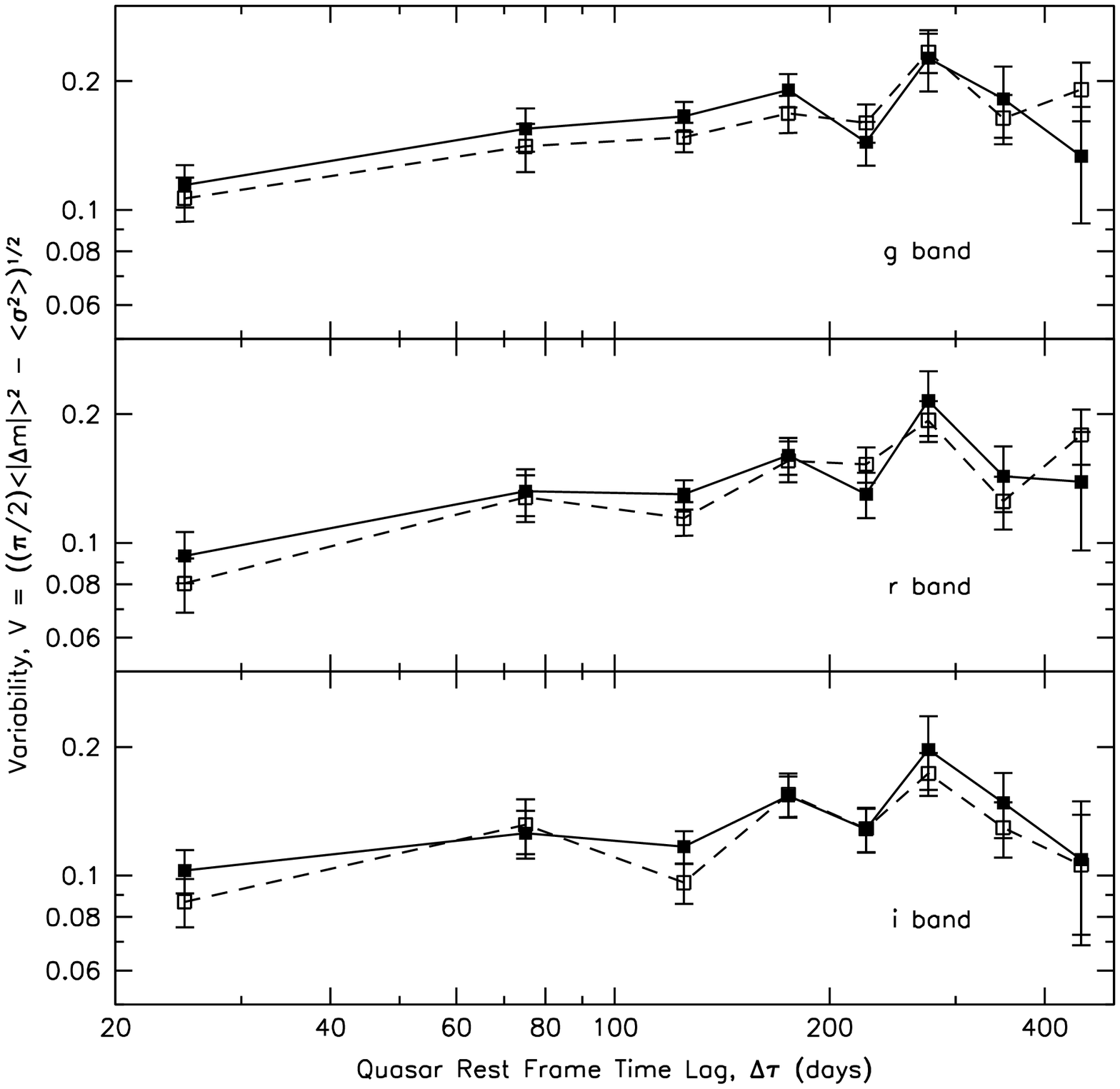}
\caption{Radio-detected (solid lines and points) vs. undetected 
  (dashed lines and open points) structure functions.
\label{Fig6.1}}
\end{figure}

\begin{figure}
\plotone{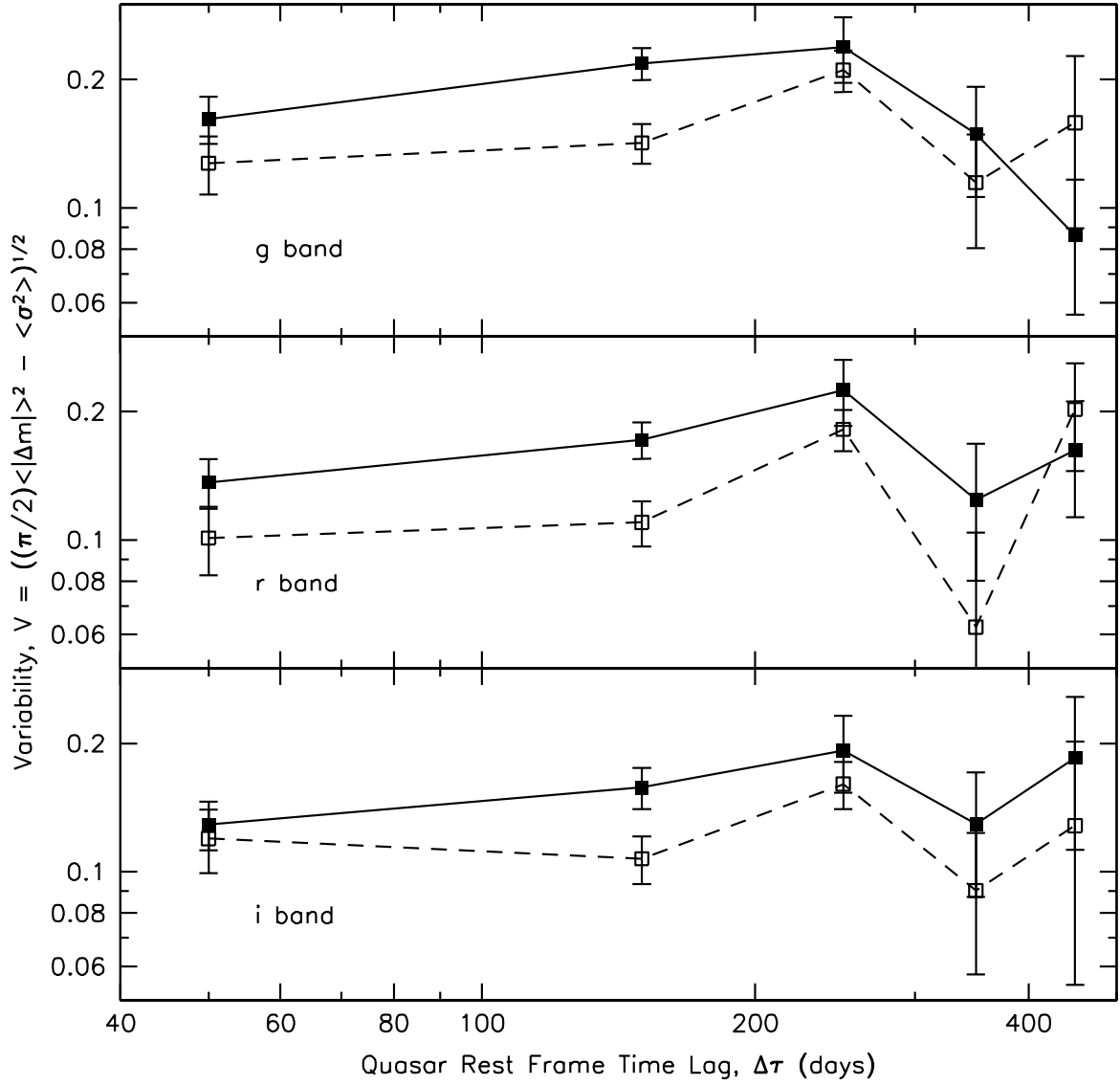}
\caption{Radio-loud (solid lines and points) vs. 
  radio-quiet (dashed lines and open points) structure functions.
\label{Fig6.2}}
\end{figure}

The matched time lag structure functions for the full radio and
non-radio samples are shown in Fig.\,\ref{Fig6.1}.  There is no
significant difference in the binned structure functions of the
radio-detected and undetected quasars.  On the other hand, the
matched structure functions for the radio-loud sub-sample, shown in
Fig.\,\ref{Fig6.2}, are about 1.3 times higher than the non-radio
sample.  However, only the difference in the $i$ band is significant
(matched pair $t$-test probability of $1\%$ if the samples were not
truly different).  Thus, there is marginal evidence that radio-loud
quasars are more optically variable than radio-quiet quasars, but a
larger sample will be needed to confirm this.

The qualitative result that the radio-loud quasars are more
variable agrees with most of the past suggestions
\citep{helfand01,eggers00,garcia99,pica83}.  Since most blazars
are radio-loud \citep[e.g.][]{ulrich97}, the higher variability
amplitude of radio-loud quasars may reflect a higher fraction
of blazars.  There is not enough information from this survey to
reliably classify individual objects as blazars (at the very least,
more detailed lightcurves are needed).  Further subdivision of the
sample by finer radio loudness currently yields too few quasars for
meaningful comparisons.  In any case, the evidence for
a correlation between radio loudness and UV/optical variability
amplitude is suggestive, but not conclusive.

%
% X-Ray
%
\subsection{X-Ray Detected Quasars\label{xray}}

%TODO: check if the X-ray sample can be sub-divided into XB bins
%  This would require extracting the X-ray flux for each object.

As with radio quasars, some of the SDSS quasars are selected for
spectroscopic follow-up as matches to sources in the Rosat All Sky
Survey \citep[RASS,][]{voges99,voges00}.  A detailed analysis of
RASS source matches to the SDSS data is given by \citet{anderson03}.
About 5\% of the verified quasars can be matched with RASS sources,
giving about 1300 X-ray quasars in our sample.  We constructed the
X-ray and control samples in the same way as for the radio sample
and its control.  The numbers of matched objects in each of the
$g$, $r$, and $i$ bands are 1010, 1008, and 1009 respectively.
Again, Kolmogorov-Smirnov tests show that with respect to
redshift, luminosity, and time lag, the two sets of samples are
indistinguishable.

\begin{figure}
\plotone{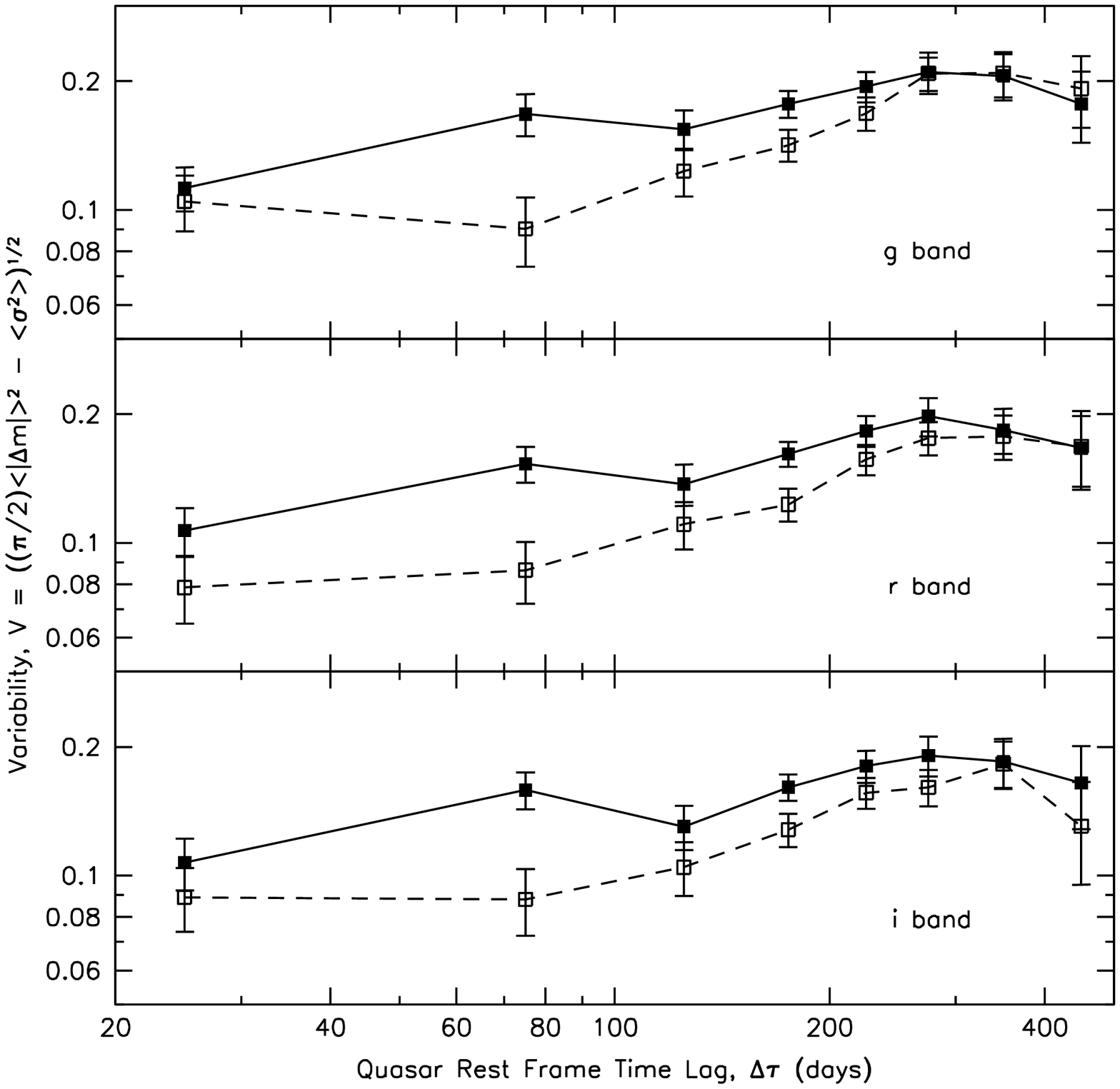}
\caption{X-ray-detected (solid lines and points) vs. 
  undetected (dashed lines and open points) structure functions.
\label{Fig6.3}}
\end{figure}

The time lag structure functions for the X-ray detected and
X-ray-non-detected sample are shown in Fig.\,\ref{Fig6.3}.  The X-ray
sample is more variable than the non-X-ray control sample at time
lags up to about 250 days, after which the differences of variability
amplitudes are much smaller.  Overall the X-ray sample amplitudes
are larger by a factor of $\approx 10\%$.  The matched pair $t$-test
probabilities of the differences occurring by chance are $9.1\%$,
$0.8\%$, and $0.4\%$ for the $g$, $r$, and $i$ bands respectively.
The data therefore show that X-ray selected quasars
are significantly more variable than their X-ray-non-detected
counterparts.  However, the difference may become less significant
at longer time lags and at longer wavelengths.

The higher X-ray variability amplitude is probably not surprising
given the high fraction of X-ray detected objects among blazars.
As for the radio sample, the X-ray sample was selected purely
by optical matches to catalog sources; no information about the
variability of the objects in the sample was used beforehand.
This is the first time that a large X-ray selected sample of quasars
has been examined for UV/optical variability.  The SDSS sample will
soon be large enough to subdivide it further by X-ray brightness.
In the meantime, a correlation between X-ray emission and UV/optical
variability amplitude can be claimed.  We discuss this further
in \S\,\ref{discussion}.

%
% BAL
%
\subsection{Broad Absorption Line Quasars\label{bal}}

Broad absorption line quasars (BALQSOs) are defined by the presence
of very strong, blue-shifted absorption troughs in their spectra.
About $5\%$ of the quasars in the SDSS sample can be classified as
BALQSOs, a fraction which is heavily redshift and color dependent.
The largest systematically selected samples of BALQSOs are those
of \citet{reichard03} and \citet{tolea02}, each of which contain
close to the same sets of objects drawn (in somewhat different ways)
from the SDSS Early Data Release quasar sample.  Both sets contain
about 200 objects, and for the present purposes the samples are
indistinguishable; we have used the \citet{reichard03} sample since
the selection process is more automated and is likely to be used
for future BALQSO catalogs.  A control sample of non-BAL quasars
was designed to have consistent redshift, luminosity, and time lag
distributions, in the manner described above.  The matched sample of
BALQSOs contains 178, 189, and 190 objects in the $g$, $r$, and
$i$ bands respectively.

\begin{figure}
\plotone{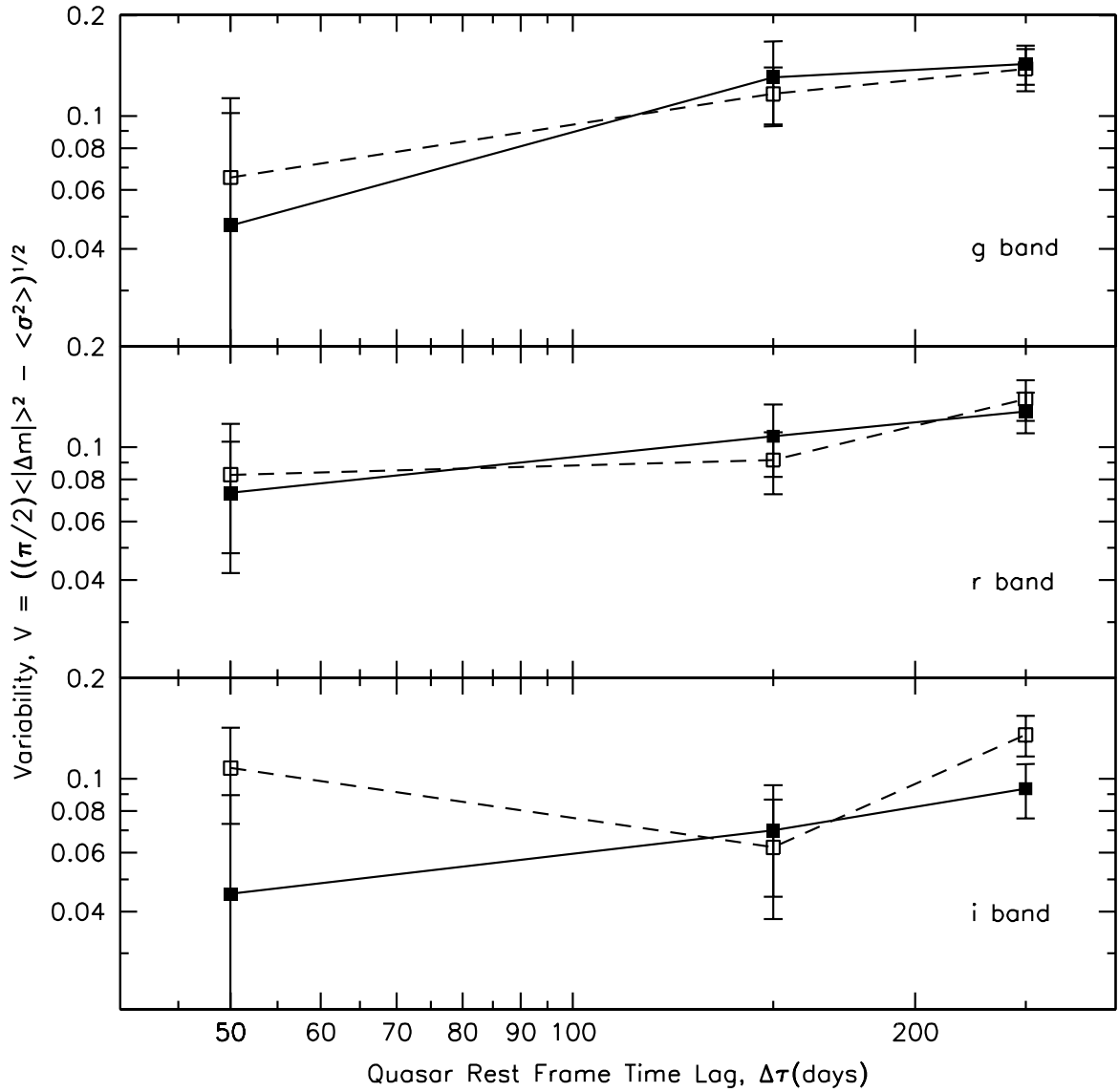}
\caption{BALQSO (solid lines and points) vs. non-BALQSO 
  (dashed lines and open points) structure functions.
\label{Fig6.4}}
\end{figure}

The matched structure functions are shown in Fig.\,\ref{Fig6.4}.  The
time lag binning is necessarily coarse due to the relatively small
sample sizes.  At the level of sensitivity of this sample, there
is no difference in the variability amplitudes of BAL and non-BAL
quasars.
%
%Clearly a larger sample is desirable, however,
%considering the full time lag range, a significant difference
%could be detected if one sample were a factor of $XXX$ larger than
%%the other.  The differences described above for the radio-loud and X-ray
%samples would be significant at the $XXX$ standard deviation level.
%
Currently favored models of the BAL phenomena attribute the absorption
to high opacity gas, either as clouds or flows, usually viewed near the
plane of an accretion disk.  If the presence of BALs is purely a viewing
angle effect, then continuum variability, if due to the central quasar engine
is unlikely to be correlated.  However, if variability is due to the
presence of obscuring dust of varying attenuation crossing the sightline
to a quasar, BAL quasars may be expected to be more highly variable
at optical and UV wavelengths.  The issue will need to be settled with
a larger sample, but the current results do not support any correlation
between the presence of BAL features and UV/optical variability.

%%%%%%%%%%%%%%
% Discussion %
%%%%%%%%%%%%%%
\section{Discussion\label{discussion}}
%We should discuss implications for current models of variability:
%microlensing; Poissonian processes (e.g. multiple supernovae,
%multiple star collisions); varying dust reddening and obscuration;
%accretion disk instabilities (e.g. accretion rate changes, hot spots,
%avalanches); jet-disk interactions (e.g. reprocessing of jet X-rays).

To summarize our results, we have separated the dependence
of variability on a number of parameters, and found a power
law dependence on time lag, anticorrelations with wavelength
and luminosity, and a correlation with redshift.  All of these
relationships have been parameterized.  Radio loud and X-ray quasars
also appear to be more variable than their quiet counterparts.  There is
currently no model of quasar continuum variability at optical and
UV wavelengths that addresses all of the relationships described
here, and until recently, there were virtually no quantitative
predictions.  Current models can be classified broadly into three
groups: accretion disk instabilities, discrete-event or Poissonian
processes, and gravitational microlensing.  We ignore other evidence
for or against the theories and describe how variability as an
independent phenomenon may constrain the models.

The Poissonian models postulate that quasar luminosity, or at least
a significant fraction of it, is generated by some type of multiple
discrete and independent energetic events, such as supernovae
or star collisions \citep[e.g.][]{terlevich92,torricelli00}.
The statistical superposition of the light curves of the randomly
occurring events determines the luminosity at any given time.
As discussed in \S\,\ref{absmag}, the simplest Poissonian models
predict a luminosity dependent power law slope of $\beta = 1/2$
which is inconsistent with our results.  More detailed models in
which the event rate, energy, timescale, and background contribution
are adjustable parameters can produce a wide range of slopes
\citep{cid00}, but a value of $\beta = 1/2$ is still difficult
to avoid in models invoking supernovae and their remnants as the
events \citep{paltani97,aretxaga97}.  Another apparently unavoidable
consequence of the Poissonian models is that the variable luminosity
component is not wavelength dependent \citep{cid00}, and any color
changes must be the result of a non-variable component (such as a
host galaxy), which must be red to qualitatively account for the
wavelength correlation found here and in other studies.  We have
shown that the variable source itself is wavelength dependent, and
a host galaxy component alone cannot account for the color changes.
Quantitative predictions for the power law slope of the structure
function in the starburst (supernova) model \citep{kawaguchi98}
range from $\gamma \approx 0.7 - 0.9$, which are also quite
inconsistent with the value we find ($\gamma = 0.246$).  Thus,
based on predictions for the time lag, wavelength, and luminosity
dependence of variability, current Poissonian models are inconsistent
with the observational results of this study.  It remains to be
seen if non-Poissonian processes, for example in which the events
are not independent or random, can account for the observations.

%%It is also difficult to explain why there
%%would be a variability dependence on radio or X-ray properties if
%%the descrete events model holds.

The idea that the motions of intervening matter along the geodesics
to quasars may cause flux variations (microlensing) was discussed
as early as the late 1970s \citep{chang79}, but few quantitative
predictions have been worked out with regard to typical quasar
variability.   Using the simulated microlensing light curves
of \citet{lewis93} and \citet{schneider87}, \citet{hawkins02}
generated structure functions which have a power law form with
slopes ranging from $\gamma \approx 0.23$ to $\gamma \approx 0.31$
\citep{hawkins02}, which is consistent with what we find.  However,
the slopes depend upon the unknown lens mass distribution function,
velocity distribution, and source size \citep{wyithe01}, so that
a wide range of values is possible.  There is little doubt that
microlensing of quasar images does happen, and it has likely been
detected in at least two cases, Q2237+0305 \citep[][and references
therein]{schmidt02} and Q0957+561 \citep{refsdal00}.  However in
each case, the quasar is {\em macrolensed} by a foreground galaxy,
which means that the geodesics selectively pass through regions
of relatively high density.  \citet{wyithe02} showed that the
probability of microlensing by stars among single image (not
macrolensed) sources is very small; dark matter composed of
compact objects can improve the probability, but at most only
about $10\%$ of sources are expected to be microlensed at any
given time.  In addition, since unresolved macrolensed quasars --
which appear more luminous than they really are -- are more likely
to be microlensed, the anticorrelation of variability amplitude
with luminosity is opposite to the trend that would be expected
from microlensing.  While the deflection of light by gravity
is achromatic, the wavelength dependence of variability is not
necessarily evidence against the microlensing hypothesis, as long
as quasar emitting regions are smaller and brighter with decreasing
wavelength.  It is difficult, however, to see why there would be any
dependence on the radio or X-ray properties of the quasar sample.
The amplitude of quasar variability changes with redshift, but it
is nearly as strong at low redshifts as it is at high redshifts,
which also shows that microlensing can not be the primary cause
of variability, since microlensing events should be extremely
rare at low redshift.  Finally, reverberation mapping studies
\citep{peterson01} show that the quasar broad line region varies
in response to changes in the continuum luminosity, showing that a
large fraction of variability must be intrinsic to quasars.  If some
quasar variability is due to microlensing, it will be important
to isolate it from the other sources since it has the potential to
constrain the components of dark matter and the structure of quasars.

It is widely accepted that quasar luminosity is generated through
some set of processes related to the accretion of matter from
a disk onto a supermassive black hole \citep[e.g.][]{rees84}.
It is therefore natural to consider mechanisms associated with
changes in these processes as a source of quasar variability.
Qualitatively, most schemes would tend to follow the trends we
find here, in particular the anticorrelations of variability with
wavelength and luminosity.  For example, the disk emission spectrum
of the standard optically thick geometrically thin accretion disk
model \citep[e.g.][]{shakura73} is more luminous and bluer when the
accretion rate is higher, and the relative luminosity change would
be lower in more luminous objects for a given change in accretion
rate.  However, it is not known how the accretion rate would change
nor how the resultant luminosity changes would propagate through
the disk.  While there has been much theoretical discussion of
the various possibilities for the emission mechanisms and their
instabilities \citep[e.g.][]{wallinder92,schramkowski96}, there
are currently few quantitative predictions which can be compared
to observational results.  \citet{kawaguchi98} generated structure
functions for the cellular-automaton model for disk instability,
and found power law forms with slopes ranging from $\gamma = 0.41$
to $0.49$.  While this range is inconsistent with our results, the
model is necessarily simplified and a number of assumptions need
to be made.  The complexity of possible accretion disk (or jet)
instability models is likely what has prevented more quantitative
predictions.  Disk instability models are clearly promising, but
as yet it is difficult to compare them to the observations.

In summary, the weight of the observational evidence seems to disfavor
gravitational microlensing and generic Poissonian processes as the
primary source of quasar variability.  Accretion disk instability
models have yet to be adequately developed quantitatively for direct
comparison with our results.  It is also plausible that a combination
of sources produce variations in quasar lightcurves, and no single
model can be completely eliminated at this time.

%%%%%%%%%%%%%%%
% Conclusions %
%%%%%%%%%%%%%%%
\section{Conclusions\label{conclusions}}
We have examined the ensemble broadband photometric variability
of a very large and homogeneous sample of quasars from the SDSS --
the largest sample ever used to study variability.  The three-band
spectrophotometry of each object was compared directly to the
imaging photometry obtained at an earlier epoch.  Because of the
large number of objects and wide coverage of parameter space,
the dependences of variability amplitude on time lag, luminosity,
wavelength, and redshift were able to be disentangled for the
first time.  The variability amplitude increases with time lag
(up to about two years) as a power-law with a slope of $\gamma
= 0.25$.  In terms of the variability amplitude, more luminous
quasars are less variable, shorter wavelengths are more variable,
and more distant quasars are somewhat more variable; all of these
relationships are parameterized.  Radio loud quasars appear to be
more variable than their radio quiet counterparts, and quasars with
detectable X-ray emission (in the ROSAT survey) are more variable
than those without.  It is difficult to explain the results in the
context of models involving discrete events (Poissonian models)
and gravitational microlensing.  Accretion disk instability models
are promising, but more quantitative predictions are needed to test
them against the observational results.

%%%%%%%%%%%%%%%%%%%
% Acknowledgments %
%%%%%%%%%%%%%%%%%%%
\acknowledgments

Funding for the creation and distribution of the SDSS Archive has
been provided by the Alfred P. Sloan Foundation, the Participating
Institutions, the National Aeronautics and Space Administration,
the National Science Foundation, the U.S. Department of Energy,
the Japanese Monbukagakusho, and the Max Planck Society. The SDSS
Web site is {\tt http://www.sdss.org/}.

The SDSS is managed by the Astrophysical Research Consortium (ARC)
for the Participating Institutions. The Participating Institutions
are The University of Chicago, Fermilab, the Institute for Advanced
Study, the Japan Participation Group, The Johns Hopkins University,
Los Alamos National Laboratory, the Max-Planck-Institute for
Astronomy (MPIA), the Max-Planck-Institute for Astrophysics (MPA),
New Mexico State University, University of Pittsburgh, Princeton
University, the United States Naval Observatory, and the University
of Washington.

%%%%%%%%%%%%%%
% References %
%%%%%%%%%%%%%%

\onecolumn
\clearpage
%%%%%%%%%%%
% Figures %
%%%%%%%%%%%
%
% Example:
% \begin{figure}
% \epsscale{0.5}
% \plotone{image.ps}
% \caption{ 
% \label{image1}}
% \end{figure}

\clearpage

\clearpage

\clearpage

\clearpage

\clearpage

\clearpage

\clearpage

\clearpage

%% Section 4 Figures
%%
\clearpage

\clearpage

\clearpage

\clearpage

\clearpage

\clearpage

\clearpage

\clearpage

%% Sec 5 Figures
%%
\clearpage

%% Section 6 Figures
%%
\clearpage

\clearpage

\clearpage

\clearpage

%%
%% Tables
%%
\clearpage
%
% Table 1. -- Parameter fits to binned structure functions
%% Parameter fits to the uncorrected structure functions
%% 
%% Columns are
%% Parameter name, fit value, 1sigma uncertainty, Chi^2
%%
\begin{deluxetable}{cccc}
\tablewidth{0pt}
%\rotate
\tablecaption{Parameter values for fits to the binned structure functions.
\label{tab3.1}}
\tablehead{
  %% upper heading
  \multicolumn{4}{c}{Power Law}
}
\tablecolumns{4}
\startdata
  Band & $\Delta\tau_{0}$ (days) & $\gamma$ & $\chi^2$ \\ \hline
 $g$ & $9.90\pm6.49 \times 10 ^{4}$ & $0.293\pm0.030$ & $1.51$  \\
 $r$ & $7.05\pm4.23 \times 10 ^{4}$ & $0.336\pm0.033$ & $1.20$  \\
 $i$ & $1.66\pm1.32 \times 10 ^{5}$ & $0.303\pm0.035$ & $4.20$  \\
\cutinhead{Exponential}
  Band & $V_{0}$ (mag) & $\Delta\tau_{0}$ (days) & $\chi^{2}$ \\ \hline
 $g$ & $0.168\pm0.005$ & $51.9\pm6.0$ & 20.5 \\
 $r$ & $0.155\pm0.006$ & $74.7\pm8.9$ & 24.8 \\
 $i$ & $0.139\pm0.005$ & $62.6\pm8.3$ & 39.3 \\
\enddata
\end{deluxetable}

%
% Table 2. -- Parameter fits to scaled variability time lag dependence
%% Parameter fits to the scaled variability time lag dependence
%% 
%%
\begin{deluxetable}{ccc}
\tablewidth{0pt}
%\rotate
\tablecaption{Parameter values for fits to the scaled variability time lag
  dependence.
\label{tab4.1}}
\tablehead{
  %% upper heading
  \multicolumn{3}{c}{Power Law}
}
\tablecolumns{3}
\startdata
 $\Delta\tau_{0}$ (days) & $\gamma$ & $\chi^2$ \\ \hline
 $5.36\pm1.46 \times 10 ^{5}$ & $0.246\pm0.008$ & 299.0 \\
\cutinhead{Exponential}
 $V_{0}$ & $\Delta\tau_{0}$ (days) & $\chi^{2}$ \\ \hline
 $0.144\pm0.001$ & $40.4\pm1.4$ & 443.4 \\
\enddata
\end{deluxetable}

%
% Table 3. -- Parameter fits to scaled variability abs. mag. dependence
%\input{tab4.2.tex}

%%
%% The End
%%
\end{document}